\pdfoutput=1
\documentclass[%
 reprint,
superscriptaddress,
groupedaddress,
 amsmath,amssymb,
 aps,
pre,
]{revtex4-1}
\usepackage{enumerate}
\usepackage{graphicx}% Include figure files
\usepackage{dcolumn}% Align table columns on decimal point
\usepackage{bm}% bold math
\usepackage{graphicx}% Include figure files
\usepackage{bbold} %for identity matrix
\usepackage{graphicx}% Include figure files
\usepackage[compat=1.1.0]{tikz-feynman}% Feynman diagrams
\usepackage{dcolumn}% Align table columns on decimal point
\usepackage[left=3cm,top=2.5cm,right=3cm,bottom=2.5cm]{geometry}
\usepackage{array}
\usepackage{epsfig}
\usepackage{wrapfig}
\usepackage{color}
\usepackage{soul}
\usepackage{braket}
\usepackage{verbatim}
\usepackage{mathtools}
\usepackage{mathrsfs}
\usepackage{amsfonts}
\usepackage{physics}
\usepackage[english]{babel}
\usepackage{textcomp}
\usepackage{float}
\usepackage{comment}
\usepackage[T1]{fontenc}
\excludecomment{toexclude}
\usepackage{changepage} %to move tables to the left using "adjustwidth"

\begin{document}

%\preprint{APS/123-QED}

\title{Quantum effects in an expanded Black--Scholes model}

\author{Anantya Bhatnagar}
\author{Dimitri D. Vvedensky}
\affiliation{The Blackett Laboratory, Imperial College London, London SW7 2AZ, United Kingdom}

\begin{abstract}
The limitations of the classical Black--Scholes model are examined by comparing calculated and actual historical prices of European call options on stocks from several sectors of the S\&P 500. Persistent differences between the two prices point to an expanded model proposed by Segal and Segal (1998) in which information not simultaneously observable or actionable with public information can be represented by an additional pseudo-Wiener process. A real linear combination of the original and added processes leads to a commutation relation analogous to that between {\color{black}a boson field and its canonical momentum in quantum field theory}. The resulting pricing formula for a European call option replaces the classical volatility with the norm of a complex quantity, whose imaginary part is shown to compensate for the disparity between prices obtained from the classical Black--Scholes model and actual prices of the test call options. This provides market evidence for the influence of a non-classical process on the price of a security based on non-commuting operators.
\end{abstract}

\maketitle

\section{Introduction}
\label{sec1}

The Black--Scholes--Merton model \cite{black73,merton73} (usually referred to as the Black--Scholes model) was proposed {\color{black}as} a rational basis for determining the fair price of options. The key idea behind this model is to hedge an asset, such as a stock, by buying and selling the asset and a call option on that asset in a way that eliminates risk.   For a European call option, the Black--Scholes model {\color{black}yields a pricing formula} based on five variables:~the strike price, the current stock price, the time to expiration, the risk-free interest rate, and the volatility, with the risk-free rate and volatility assumed to be constant \cite{mantegna00,wilmott06,hull15}.   Because stock prices fluctuate, the value of an option fluctuates accordingly. The Black--Scholes model assumes that stock prices follow a stationary lognormal distribution or, equivalently, that the fractional stock prices follow a normal distribution. Despite the known inadequacy of the normal distribution for market movements, the {\color{black}resulting} closed-form formulae for the price of an option at any time prior to expiry provide {\it post hoc} justification for this assumption. With many refinements and adjustments, the Black--Scholes model and its variations have become the {\it de-facto} standard for estimating the price of stock options and other financial instruments.

The mathematical form of the Black--Scholes equation has led to parallels being drawn with quantum mechanics.  As a linear second-order partial differential equation, the Black--Scholes equation can be interpreted as an imaginary-time Schr\"odinger equation, whose solution can then be formulated as a path integral \cite{dash88,dash89,linetsky98,dash04,baaquie04,kleinert04a}.  This provides a natural setting for relaxing some of the assumptions of the original Black--Scholes model, such as allowing for stochastic volatility \cite{baaquie97,lemmens08} and stochastic interest rates \cite{lemmens08}, and extending its applicability to other types of financial instruments, such as barrier and Asian options \cite{devreese10,bornetti04}.  Evaluating such path integrals benefits from the vast array of established methods for carrying out exact, approximate, and numerical evaluations \cite{dash88,dash89,linetsky98,dash04,baaquie04,kleinert04a,capuozzo21}.  Nevertheless, these studies represent an interpretation of the Black--Scholes equation based on a particular formulation of {\color{black}solutions to Schr\"odinger's equation in imaginary time}, rather than providing a link to {\color{black}the fundamental principles of quantum mechanics}.

Several groups have used quantum mechanical concepts from the outset to obtain inherently non-classical Black--Scholes models.  Broadly speaking, these studies are based either on non-commutativity (the Heisenberg picture) \cite{segal98,accardi07,melnyk08,hicks19,hicks20}, or the Black--Scholes Hamiltonian and wave function (the Schr\"odinger picture) \cite{ishio09,haven02,choustova07,choustova08,contreras10}.  Here, our focus is on the work of  Segal and Segal \cite{segal98}, who introduced quantum effects into the Black--Scholes model to incorporate market features such as the impossibility of simultaneously measuring prices and their instantaneous forward time derivatives. They argued that such effects provide a natural explanation for the extreme irregularities in the short-term movements of market prices. The mathematical framework for the Segal-Segal model is built on the formalization of non-commuting operators, which, through an accompanying uncertainty principle, is a pillar of quantum mechanics. Segal and Segal modified the Black--Scholes model by adding a new stochastic process to the Black--Scholes stochastic differential equation to account for the influence of factors not simultaneously measurable with those in the original Wiener process.  They determined a calculus for dealing with such processes and obtained a modification of the Black--Scholes pricing formula.

In this paper, we examine the efficacy of the non-classical, or `quantum' Black--Scholes model proposed by Segal and Segal \cite{segal98} for the valuation of options by considering options prices {\color{black}based on} several types of underlying stock {\color{black}in} various sectors of the S\&P 500.  We will first compute the price based on the original Black--Scholes model with estimates of the interest rate and volatility from market data.  The quantum Black--Scholes pricing formula \cite{segal98} is then used to find the magnitude of the additional volatility in order to associate discrepancies in the original calculations with the presence of the non-classical effects predicted by Segal and Segal. To our knowledge, this is the first time that the Segal--Segal model has been tested against actual market data.

The outline of our paper is as follows. In Sec.~\ref{sec3}, we briefly review the salient points of the original Black--Scholes model  \cite{black73,merton73} for European options, including the modified model with time-dependent interest rates and volatility.   The time-dependent Black--Scholes pricing formulae are used to calculate the prices of call options based on several stocks by using publicly available market data for the strike price, the current stock price, and the time to expiration, and to estimate the risk-free interest rate and the volatility.  The basic principles behind the quantum Black--Scholes model are reviewed in Sec.~\ref{sec4}. The quantum pricing formulae obtained from this model are used to calculate the imaginary contributions to the generalized (implied) volatility in order to identify the origins of discrepancies between actual and calculated option prices in Sec.~\ref{sec3}. A summary and our conclusions are provided in Sec.~\ref{sec5}. Additional comparisons between the Black--Scholes and expanded Black--Scholes models with market data for European call options are provided in the Supplementary Material.

\section{Types of option}
\label{sec2}

Options come in two main forms:~calls, which give the owner of the option the right (but not the obligation) to buy the underlying security at a fixed strike price before or on a specified date (the expiry date), depending on the type of option, and puts, which give the owner the right (but not the obligation) to sell the underlying security at a fixed strike price before or on a specified date. These are called long positions:~the investor owns the right to buy or sell to the writer of the option at the strike price. Conversely, writing a call or put option, whereby the writer must sell to or buy from the long position holder or buyer of the option, is known as holding a short position.

The most common option styles are European, which can be exercised only at expiry, and American, which can be exercised at any time prior to expiry (and so are worth more). For example, if an investor buys shares in Apple, Inc.~(AAPL) at \$200, but wants to protect the investment from the downside, that investor can buy puts or sell calls at a strike price of, say, \$200. If a short call position is taken, the investor pockets the premium if the share price is below \$200 at expiry of the option. Likewise, taking a long position on a put, the investor can buy AAPL shares at the market price at option expiry and sell using the right obtained from the put ownership, thereby making a profit. Although the maximum amount at risk for a long position is the premium paid, the maximum risk is the strike price for a short put and unlimited for a short call.

\section{Black-Scholes model}
\label{sec3}

\subsection{Black--Scholes pricing formulae}
\label{sec3.1}

This section provides a brief summary of the main steps for obtaining the Black--Scholes formulae for a European call option.  More detailed derivations may be found in Refs.~\cite{mantegna00,wilmott06,hull15}.  There are two types of assets in the Black--Scholes model:~a risky (i.e.~fluctuating) asset $S(t)$, such as a company stock, which is assumed to follow geometric Brownian motion with drift, and a risk-free asset $B(t)$, such as a bank account or a Treasury Bill {\color{black}(T-Bill)} paying an interest rate $r$.  The stock price at time {\color{black}$t \in [0,T_m]$}, where $T_m$ is the {\color{black}fixed duration between the issuance of the option and its maturity}, is
\begin{equation}
dS=\mu S\,dt+\sigma S\, dW\, ,
\label{eq1}
\end{equation}
where $\mu$ is the percentage drift,  $\sigma$ the percentage variance, both assumed constant, and $W$ is a standard Wiener process, also known as standard Brownian motion. The differential equation for $B(t)$ is
\begin{equation}
dB=rB\,dt\, ,
\end{equation}
whose solution is $B(t)=B(0)e^{rt}$, where $B_0$ is the initial balance of a bank account or the initial value of a Treasury Bill.

We consider an option on a stock with strike price $K$ and time to maturity $T{\color{black}=T_m - t}$.  As the stock fluctuates, so does the value $V(S,t)$ of the option.  The stochastic differential equation for $V(S,t)$ is, from It$\hat{\rm o}$'s lemma,
\begin{equation}
dV=\bigg(\mu S{\partial V\over\partial S} +{\partial V\over\partial t}+{\sigma^2 S^2\over2}{\partial^2V\over\partial S^2}\bigg)dt+\sigma S{\partial V\over\partial  S}\,dW\, .
\end{equation}

Consider a portfolio that contains the option, which has been sold (short position), and $\Delta$ shares of the underlying asset (long position). The value $\Pi$ of this portfolio is
\begin{equation}
\Pi=\Delta S-V(S(t),t)\, .
\label{eq4}
\end{equation}
According to It$\hat{\rm o}$'s lemma, the stochastic differential equation for $\Pi$ is
\begin{equation}
d\Pi=-\bigg({\partial V\over\partial t}+{\sigma^2 S^2\over2}{\partial^2V\over\partial S^2}\bigg)dt+\bigg(\Delta-{\partial V\over\partial S}\bigg)dS\, .
\label{eq5}
\end{equation}
Black and Scholes \cite{black73} observed that the coefficient of $d S$, which corresponds to the random part of the Brownian motion and leads to volatility and, therefore, to risk, can be eliminated by setting $\Delta = \partial V/\partial S$. The resulting differential equation (\ref{eq4}) is deterministic. Once we obtain a solution for $V$, this relation provides a hedging prescription for maintaining an instantaneously riskless portfolio.  We now invoke the assumption that there is no arbitrage in the market, so any riskless portfolio must earn the risk-free interest rate of the market:~$d\Pi=r\,\Pi\,dt$. Using this equation with (\ref{eq4}) and (\ref{eq5}) leads to the celebrated Black--Scholes equation:
\begin{equation}
{\partial V\over\partial t}+{\sigma^2S^2\over2}{\partial^2V\over\partial S^2}+rS{\partial V\over\partial S}-rV=0\, .
\end{equation}
Explicit formulae for the values of options are obtained by transforming this linear parabolic equation into the backwards heat equation (also linear and parabolic), which is solved with the appropriate boundary conditions.  For European call options $C(S,T)$, after transforming back to the original variables, we have that
\begin{gather}
C(S,0)=\mbox{max}(S-K,0)\, ,\\
C(0,T)=0\, ,\\
\lim_{S\to\infty}C(S,T)=S\, ,
\end{gather}
for {\color{black}$T_m\ge T\ge 0$.} Therefore, the solution to the Black--Scholes equation is
\begin{equation}
C(S,T)=SN(d_1)-Ke^{-rT}N(d_2)\, ,
\label{eq8}
\end{equation}
in which $N(x)$ is the cumulative normal distribution function, and
\begin{gather}
\begin{aligned}
d_1&={\ln(S/K)+\big(r+{1\over2}\sigma^2\big)T\over\sigma\sqrt{T}}\, ,\\
d_2&={\ln(S/K)+\big(r-{1\over2}\sigma^2\big)T\over\sigma\sqrt{T}}\, .
\label{eq9}
\end{aligned}
\end{gather}

\subsection{Stock option pricing from market data}
\label{sec3.2}

The pricing formulae (\ref{eq8}) and (\ref{eq9}) for a European call option are specified by five variables:~the strike price $K$, the current stock price $S$, the time $T$ to expiration, the risk-free interest rate $r$, and the volatility $\sigma$. The strike price and the time to expiration are set by the writer of the option.  The current stock price is available from several sources to whatever {\color{black}frequency} desired.  We have used daily stock price returns.  The specification of interest rates and the volatility merit further discussion.  For the interest rate, we take the continuously compounded yield on a 3-month T-bill whose maturity date is closest to the expiry date of the option. T-bills are guaranteed by the government of the United States and are, therefore, considered to be free of default risk. Because the interest rates were so low, and the changes were essentially negligible over the 6-week period of our options window, we took the constant interest rate of 0.08\% \cite{cnbc} for all our calculations.

An altogether different approach is used for the instantaneous volatility of a stock.  We consider companies listed on the S\&P 500, which is an index determined by the 500 largest companies in the United States listed on the New York Stock Exchange or NASDAQ ranked and weighted by total market capitalization (the product of the share price and the number of shares of a corporation held by stock holders). We determine the percentage variance for each stock from
\begin{equation}
\sigma=\beta{\mbox{VIX}\over100}\, .
\label{eq10}
\end{equation}
Here, VIX is the ticker symbol for the volatility index of the Chicago Board Options Exchange, which is a real-time measure of volatility based on S\&P 500 index options with near-term expiration dates \cite{fleming95,vix}.  VIX is a {\it forward-looking index} obtained from the implied volatilities of S\&P index options and, therefore, represents the market's expectation of the 30-day future volatility of the S\&P index, while incorporating the information content of historical (i.e. backward-looking) volatility. The factor $\beta$ is a {\it backward-looking} measure of the volatility of a stock compared to the volatility of all other stocks in a particular index, in our case, the S\&P 500 index \cite{beta}.  The regimes of interest are: $0<\beta<1$ for stocks with a lower volatility than the S\&P 500, $\beta=1$ for stocks with the same volatility, and $1<\beta$ for stocks with a higher volatility than this index. The value $\beta=0$ means the stock is uncorrelated with the S\&P 500, and $\beta<0$ that the stock is negatively correlated. For the calculations reported here, we used the following (constant) values of $\beta$ \cite{beta}:
\begin{gather}
\begin{aligned}
\mbox{AAL}&=1.71\, ,&\qquad \mbox{BRK-B}&=0.84\, ,\\
\mbox{JPM}&=1.12\, ,&\qquad \mbox{NKE}&=0.82\, ,\\
\mbox{RCL}&=2.76\, ,&\qquad\mbox{TSLA}&=1.97\, .
\end{aligned}
\end{gather}
for the American Airlines Group (AAL), Berkshire Hathaway Class B (BRK-B), (c) J.~P.~Morgan Chase \& Co (JPM), Nike Inc. (NKE), Royal Caribbean Cruises Ltd, and Tesla Inc. (TSLA). These companies represent three of the eleven sectors of the S\&P 500:~industrials (AAL), financials (BRK-B and JPM), and consumer discretionary (NKE, RCL, and TSLA).

In the Black--Scholes pricing formulae (\ref{eq8}) and (\ref{eq9}), the interest rate and volatility are constant.  Allowing the interest rates and the volatility to be time-dependent (but not stochastic) leads to modified pricing formulae \cite{wilmott06}:
\begin{align}
C^\prime(S,T)&=SN(d_1^\prime)\nonumber\\
&\quad-K\exp\bigg(-\int_0^T r(\tau)\,d\tau\bigg)N(d_2^\prime)\, ,
\label{eq11a}
\end{align}
where the new functions $d_1^\prime$ and $d_2^\prime$ are now given by
\begin{align}
d_1^\prime&={\ln(S/K)+\int_0^T r(\tau)\,d\tau+{1\over2}\int_0^T\sigma^2(\tau)\,d\tau\over\sqrt{\int_0^T\sigma^2(\tau)\,d\tau}}\, ,\\
d_2^\prime&={\ln(S/K)+\int_0^T r(\tau)\,d\tau-{1\over2}\int_0^T\sigma^2(\tau)\,d\tau\over\sqrt{\int_0^T\sigma^2(\tau)\,d\tau}}\, .
\label{eq12a}
\end{align}
As we are taking the interest rate as constant, the factors in these pricing formulas that include the interest rate reduce to those in (\ref{eq9}).
Prices of call options were calculated from time series for the volatility. For each time $T$, this necessitated replacing the integrals in (\ref{eq12a}) with the corresponding Riemann sums.  Historical data \cite{bloomberg} was used for the closing price of the call option on each trading day over the 6-week period from October 8, 2020 to November 20, 2020. VIX was obtained from Yahoo Finance \cite{yahoo} and $\beta$ from CNBC \cite{cnbc}.  

%%%%%%%%%%%%%%%%%%%%%%%%%%%%%%%%%%%%%%%%%%%%%%%%%

\begin{figure*}[t!]
\centering
\includegraphics[width=0.8\textwidth]{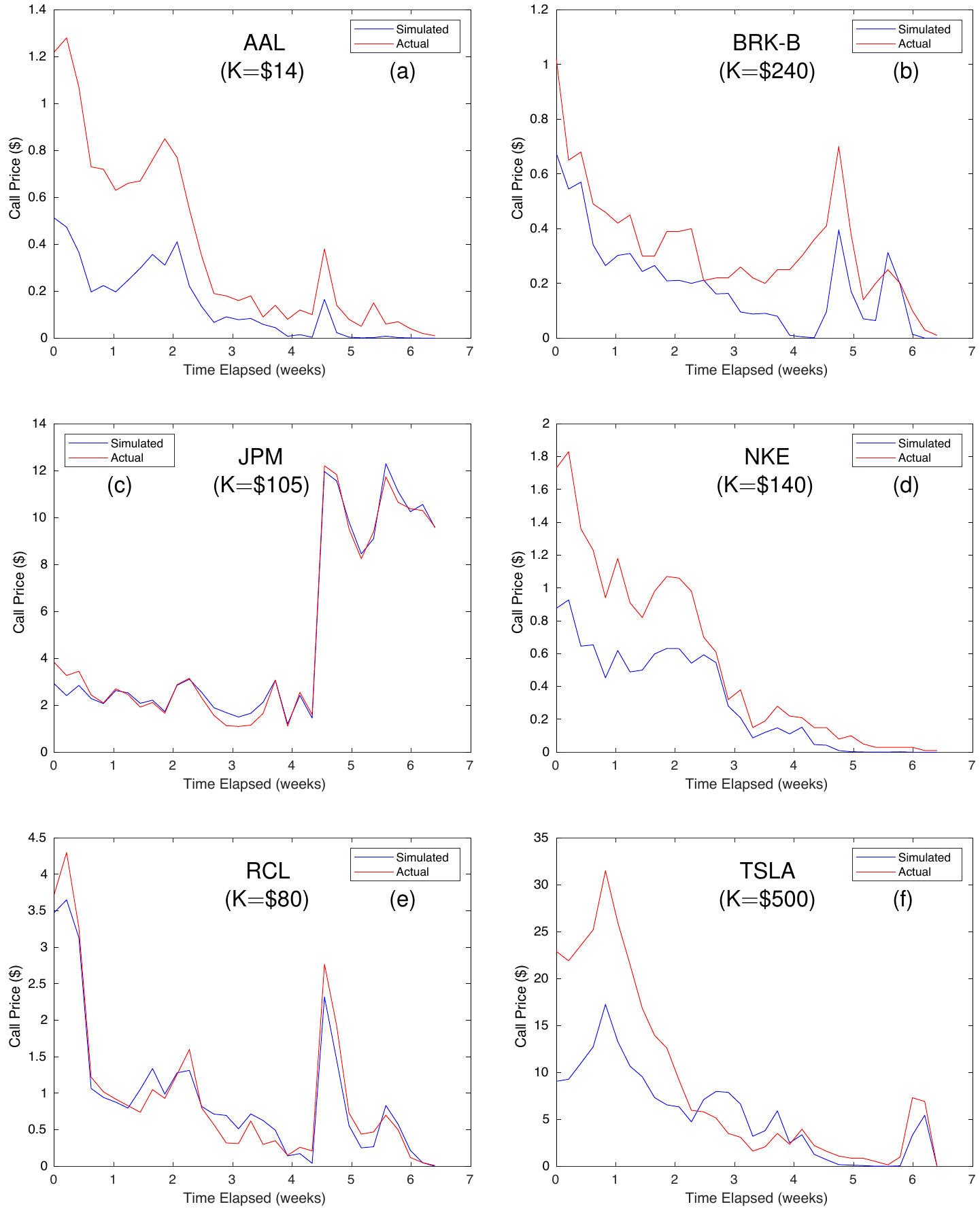}
\caption{Comparison between actual prices of European call options (red) and prices calculated from the pricing formulas (\ref{eq8}) and (\ref{eq9}) (blue) with parameters as described in the text and the indicated strike price for (a) AAL, (b) BRK-B, (c) JPM, (d) NKE, (e) RCL, and (f) TSLA. The actual prices were taken over the lifetime of the options, i.e.~the 6 weeks from the issuance date, October 8, 2020, to the maturity date, November 20, 2020, with the horizontal axis representing the time elapsed $t$ since the issuance date. The time to maturity, as used in (\ref{eq8}) and (\ref{eq9}), is then $T = T_m - t$, where $T_m$ is the fixed duration between the issuance and maturity dates.}
\label{fig1}
\end{figure*}

%%%%%%%%%%%%%%%%%%%%%%%%%%%%%%%%%%%%%%%%%%%%%%%%%

Figure~\ref{fig1} shows the comparison between the actual price of a European call option and the prices calculated from the pricing formulae (\ref{eq11a}) and (\ref{eq12a}) with the interest rates and percentage volatilities as described above.  There are several noteworthy trends in these comparisons.  Most apparent is that the Black-Scholes prices reproduce the gross qualitative trends of the actual prices for options based on each of the stocks in Fig.~\ref{fig1}.  Looking more closely shows that the Black--Scholes price is typically lower than the actual prices, with the two prices sometimes showing substantial differences.  Nevertheless, there are fleeting moments where the Black--Scholes price exceeds the actual price. 

The degree to which the Black--Scholes model reproduces the actual prices depends on the stock used for the option.  For example, the price of an option based on stock of the American Airlines Group (Fig.~\ref{fig1}(a)) is consistently underestimated by the Black--Scholes model, while options based on J.~P.~Morgan Chase \& Co (Fig.~\ref{fig1}(c)) and Royal Caribbean Cruises (Fig.~\ref{fig1}(e)) are reproduced quite well by the Black--Scholes price over the entire 6-week period. However, the Black--Scholes model shows large discrepancies for options based on stock from Berkshire Hathaway Class B, Nike, and Tesla (Figs.~\ref{fig1}(b,d,f)). Comparisons between actual and calculated option prices based on other stocks may be found in the Supplementary Material. 

While it is possible in principle to extend our analysis to intra-day resolutions, this is unlikely to yield significant improvements, due primarily to the behavior of participants in the options market. Options tend to be traded much less often compared to their underlying stocks, so intra-day price histories tend to be incomplete or unequally spaced in time, with the extreme case being tick data. This means that standardized comparisons between various options lose their noteworthiness.  The main conclusion is that, while the agreement in the broad trends of the price trajectories (and, in some cases, the accuracy of the prices) provides validation for our method, the discrepancies also point towards the possibility of factors that are not taken into account by the usual Wiener process in (\ref{eq1}).

\begin{figure*}[t!]
\centering
\includegraphics[width=0.8\textwidth]{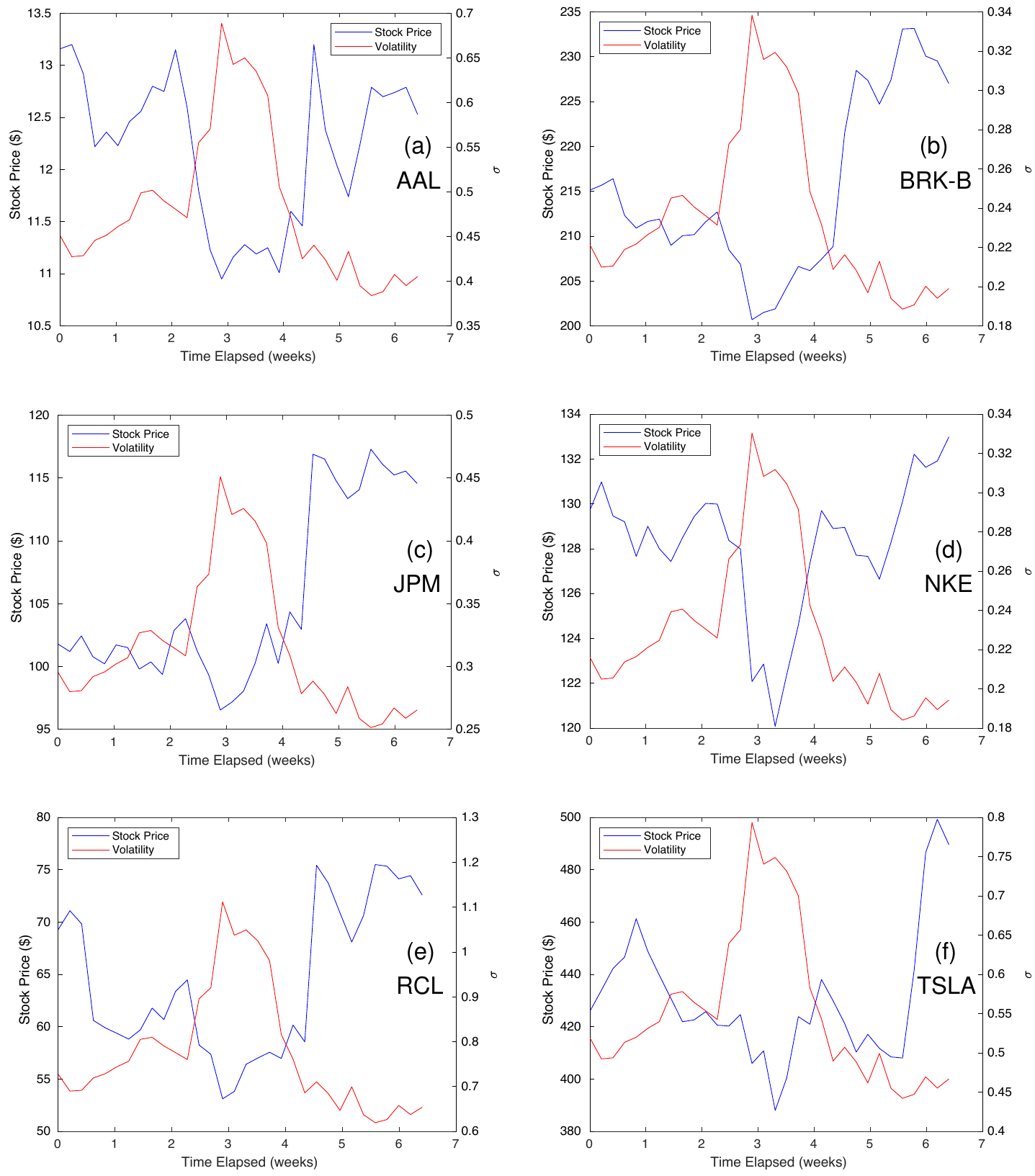}
\caption{The variations of the stock prices (blue lines) and volatilities (red lines) over the same 6-week period for the same call options with the same strike prices as in Fig.~\ref{fig1}.  The left and right vertical scales are for the stock prices and volatilities, respectively.}
\label{fig2}
\end{figure*}

Figure~\ref{fig2} shows the variations of stock prices and volatilities for the same options as in Fig.~\ref{fig1}.  Most apparent from this figure is the sharp drop in stock prices with a concomitant increase in the volatility midway between weeks 2 and 3, followed by a reversal between weeks 4 and 5.  The first period corresponds to the time when the S\&P500 fell by some 200 points due to the rise in coronavirus cases, the reintroduction of lockdowns and the lack of fiscal stimulus, compounded by the uncertainty of the presidential election in the United States. The VIX generally trades in opposite directions to stocks so the opposite movement is in line with the expected behavior (more fear amongst investors manifesting as increased volatility). Options tend to be more forward-looking than stocks, so the market sentiment reflected a more positive outlook over the longer term, given the earnings beats that were reported in the third quarter of 2020.

A comparison of Figs.~\ref{fig1} and \ref{fig2} shows that the options react in substantially different ways to the variations in the underlying stock.  Options based on stocks in the financials sector (BRK-B and JPM) showed marked increases in prices that coincide with the corresponding 13\% and 17\%  increases, respectively, in their stock prices.  In the consumer discretionary sector (NKE, RCL, and TSLA), only RCL showed evidence of the steep increase;~the stock increase was near 20\%.  The TSLA and NKE options showed no unusual behavior, which is likely due to increases of less than 10\%.

%%%%%%%%%%%%%%%%%%%%%%%%%%%%%%%%%%%%%%%%%%%%%%%%%

%%%%%%%%%%%%%%%%%%%%%%%%%%%%%%%%%%%%%%%%%%%%%%%%%

\section{Quantum Black--Scholes model}
\label{sec4}

\subsection{Volatility in the classical Black--Scholes model}

The comparisons in Fig.~\ref{fig1} used the `classical' Black--Scholes model \cite{black73,merton73} of a European call option based on publicly available information.  These comparisons reveal several {\color{black}types of trend}: 

\begin{enumerate}

\item[(i)] The two prices track each other, with the simulated price below the actual price and a time-varying difference that may approach 50\%, but with no crossing.  Figures~\ref{fig1}(a,d) show this type of behavior.  

\item[(ii)] Same as (i), but with crossing.  Figures~\ref{fig1}(b,e,f) show this type of behavior.

\item[(iii)] The two prices track each other, with a small time-varying difference that does not exceed a few percent, and with crossing.  Figure~\ref{fig1}(c) shows this type of behavior.  

\item[(iv)] The simulated price exceeds the actual price for most or all times during the lifetime of the call. Examples are provided in the Supplementary Material.

\end{enumerate}

Figure~\ref{fig2} shows that there are large swings in the volatility with proportionately smaller swings in the prices of the underlying stocks. Of the five quantities that determine the formulae (\ref{eq11a}) and (\ref{eq12a}), the volatility of the stock is the most difficult to estimate because, unlike the other four parameters, volatility cannot be observed directly and, therefore, {\color{black}has been} open to interpretation.  According to the pricing formulae (\ref{eq11a}) and (\ref{eq12a}), the price of a European call option increases with the volatility.  Thus, an accurate estimate of the price of an option depends on an accurate volatility.  The most common estimates of the volatility are based on the historical volatility of the stock (backward looking) and the implied volatility of the option (forward looking) \cite{mantegna00,wilmott06,hull15}.

Historical volatility depends on past performance of the stock and, in this regard, is not necessarily reflective of future volatility. This gives it limited use in the calculation of option prices, which by their very nature are inherently forward-looking. Implied volatility involves using the prices of near-the-money options, which are largely set by market forces of supply and demand, to calculate the resulting volatility using the Black-Scholes formulae \eqref{eq8} and \eqref{eq9}. This implies that one assumes the Black-Scholes price to fully reflect the market price, even taking into account the assumptions that are known to make it an idealized model, rather than a true reflection of market dynamics. 

We propose VIX as a better estimate for the following reasons. Firstly, its calculation does not depend on the Black-Scholes formula, but instead is model-independent, and is found using  \cite{vix21,derman99}
\begin{equation}
    F = \textrm{Strike} + e^{RT} (\textrm{Call}\,\textrm{Price} - \textrm{Put}\,\textrm{Price})\, ,
\end{equation}
from which we obtain
\begin{equation}
    \sigma^2 = \frac{2}{T}\sum_i \frac{\Delta K_i}{K_i^2}e^{RT}Q(K_i) - \frac{1}{T}\left[\frac{F}{K_0} - 1\right]^2\, ,
    \end{equation}
and
\begin{equation}
    \textrm{VIX} = 100 \times \sqrt{\sigma^2}\, .
    \label{eq24}
\end{equation}
Here, $T$ is the time to expiration, $F$ is the forward index level derived from index option prices (i.e.~determined by identifying the strike price at which the absolute difference between the call and put prices is the smallest), $K_0$ is the first strike below the forward index level $F$, $K_i$ is the strike price of the $i^{th}$ out of the money option, $\Delta K_i = {1\over2}(K_{i+1} - K_{i-1})$ is the interval between the strike prices, $R$ is the risk-free interest rate and $Q(K_i)$ is the midpoint of the bid-ask spread for each option with strike $K_i$.

Thus, the calculation of VIX is unambiguously formulated and (more significantly) not restricted by the usual Black-Scholes assumptions, making it more reflective of market conditions than an implied volatility based on the latter. Secondly, the formula \eqref{eq24} uses the prices of 30-day forward index options, and so is forward-looking. This makes it a more reliable indicator than historical volatility, since it is a measure of the future expected volatility of the index.

The central conclusion of Sec.~\ref{sec3} is that option prices determined by (\ref{eq11a}) and (\ref{eq12a}) are typically below the actual price, with exceptions resulting from particular events that cause pessimism amongst investors.  The particular events and the effect on the stock and option prices are best explained on a case-by-case basis.  The reason that such events lead to calculated option prices exceeding actual prices is that our volatility, which is obtained from and reflects public information, may not change quickly enough to account for the reaction of investors to such news (over small time scales).  Indeed, there are several instances in Fig.~\ref{fig1} where the trends in calculated and actual prices differ significantly. In general, a falling stock price implies higher volatility (due to added fear) (Fig.~\ref{fig2}), but a greater volatility in (\ref{eq11a}) and (\ref{eq12a}), with all other quantities held constant, yields a higher call price.  Hence, we must look to the interplay between the changes in stock price and volatility, and which has a larger effect on the option price. In the resolution of this, the approach of Segal and Segal  \cite{segal98} provides a viable alternative to the classical Black--Scholes model.

{\color{black}One may ask whether it is valid to simply attribute a discrepancy between market volatility and that obtained from any model to an inaccurate estimation (i.e. the inevitable experimental error), and take the market price as reflecting the true volatility $\sigma$ of a given stock. However, this does not take into account the fact that the information content of volatility is derived from multiple sources, all of which may not be simultaneously actionable independent of the observer. In particular, we note that non-public information here holds a different meaning to the usual connotation of additional information that might be known to, and thus simultaneously usable by, a company insider (which merely increases the existing information content of the classical volatility $\sigma$). Hence, while using VIX does correspond to an accurate estimate of the usual parameter $\sigma$, simply taking a different value does not adequately reflect a stock's true volatility. In essence, we bring into focus the limit to which the meaning of the usual volatility is applicable. As we show in the next section, considering public and non-public information (for example, the prices of stocks and their instantaneous forward time derivatives respectively) as non-simultaneous observables in the physical sense implied by quantum field theory (i.e. a nontrivial equal time commutation relation) captures this aspect well, yielding a generalized volatility parameter.}

\subsection{Quantum Black--Scholes model and pricing formulae}

Segal and Segal \cite{segal98} introduced non-classical effects into the Black--Scholes model as a way of incorporating market features, such as the impossibility of simultaneous measurement of prices and their instantaneous derivatives. The basic idea is to add to the Wiener process $W$ in (\ref{eq1}) for the evolution of public information, a process $X$ that represents processes that are not simultaneously observable with those in $W$. By invoking the formal structure of {\color{black}bosonic} quantum field theory \cite{jaffe05}, Segal and Segal showed that the linear combination $aW+bX$ can be represented as $\Phi((a+ib)c_t) = \Phi(f(t))$, where $\Phi$ is a mapping from vectors in a {\color{black}complex} Hilbert space $\cal{H}$ to Hermitian operators in {\color{black}the quantum field} Hilbert space $\cal{K}$ and $c_t$ is the characteristic function of the interval $[0,t]$. {\color{black}The extension from real to complex functions means that the generalized process is referred to as a pseudo-Wiener process.} For any element $f$ of a Hilbert space $\cal{H}$, $e^{i\Phi(f)}$ is the corresponding Weyl operator, whose definition is restricted to the interval $[0,t]$ because of the characteristic function. Hence, the basic equation (\ref{eq1}) becomes
\begin{align}
dS&=\mu S\,dt+\sigma S\,dW+bS\,dX\\
&=\mu S\,dt+S\,d\Phi(f(t))\, .
\label{eq14a}
\end{align}

To understand the significance of the mapping $\Phi$, we invoke the duality transform between the particle and wave representations of a quantum particle (Theorem 3 in Ref.~\cite{segal56}). The fundamental connection to the principles of quantum mechanics is readily established, where the operators representing public and private information are generalizations \cite{segal63} of the mappings in Refs.~\cite{wiener39,kakutani50}. Applying the duality transform to $e^{i\Phi(f)}$ (Theorem 4 in Ref.~\cite{segal56}), we obtain, for arbitrary $f,g$ in a real Hilbert space, the usual canonical commutation (Weyl) relations (Corollary 4.1 in Ref.~\cite{segal56}), which take the form
\begin{equation}
	e^{i\Phi(\textit{f})}e^{i\Phi(\textit{g})} = e^{i\Phi(\textit{f}+\textit{g})}e^{{1\over2}i\hskip0.5pt\textrm{Im}(\langle f,g \rangle)}
	\label{eq13}
\end{equation}
in a complex Hilbert space {\color{black}$\cal{H}$} \cite{segal61}.

The operators $\Phi(f)$ mutually commute if $f\in\mathbb{R}$ in $\cal{H}$, and likewise for $\Phi(if)$.  However,
\begin{equation}
\big[\Phi(f),\Phi(ig)\big]=i\langle f,g\rangle\, ,
\label{eq14}
\end{equation}
for $f,g\in\mathbb{R}$, where the right-hand side is an inner product defined on ${\cal H}$. This is shown to be formally equivalent to \eqref{eq13} by taking the closure of the conventional boson field creation and annihilation operators and using the appropriate infinitesimal form of \eqref{eq13}, 
\begin{equation}
    \big[\Phi(f),\Phi(ig)\big] = -i\,\textrm{Im}\big(\langle f,ig\rangle\big)\,,
\end{equation}
as in \cite{segal61}. The non-vanishing commutator \eqref{eq14} makes apparent the underlying reason why the combination of the public and private information processes cannot be modelled as two-dimensional Brownian motion.  

Returning to (\ref{eq14a}), the real part of $f(t)$ can be regarded as the process representing public information, whereas the imaginary part corresponds to a process that cannot be observed simultaneously with public information {\color{black}(for example, the price of a stock and its instantaneous forward time derivative respectively)}. The Feynman--Kac formula can then be extended in the non-commutative form (see \cite{segal98,arveson83,duffie88} for a detailed derivation) and the normal distribution of $\Phi(x)$ used, with 
\begin{equation}
\mu=0\,  ,\qquad \sigma^2=\textstyle{1\over2}\gamma^2|x|^2\, ,
\end{equation}
so that the stock price follows a lognormal distribution over $[0,T]$ with
\begin{equation}
\mu=T\big(r-\textstyle{1\over2}k\big)\, ,\qquad \sigma^2={T\over2}\gamma^2|f(T)|^2\, .
\end{equation}
The price of a call option is then expressed as
\begin{equation}
C^{\prime\prime}(S,T)=S e^{{1\over2}T(\gamma^2s^2-k)}N(d_1^{\prime\prime})-Ke^{-rT}N(d_2^{\prime\prime})\, ,
\label{eq17}
\end{equation}
where
\begin{align}
d_1^{\prime\prime}&={\ln(S/K)+\big(r+\gamma^2s^2-{1\over2}k\big)T\over\gamma s\sqrt{T}}\, ,\\
d_2^{\prime\prime}&={\ln(S/K)+\big(r-{1\over2}k\big)T\over\gamma s\sqrt{T}}\, .
\label{eq18}
\end{align}
Here, $s=|f(T)|$, $\gamma\ge1$, and $k=s^2$ or $k=0$ for Wiener or serial correlated processes, respectively. Comparing the denominators for this pricing formula with the corresponding denominators in (\ref{eq8}) and (\ref{eq9}) for the classical Black--Scholes model suggests that $\gamma s$ generalizes the classical volatility $\sigma$.  Indeed, by setting $\gamma=1$, $s=\sigma$, and $k=\sigma^2$ in (\ref{eq17}) and (\ref{eq18}), we recover (\ref{eq8}) and (\ref{eq9}).  Finally, (\ref{eq17}) and (\ref{eq18}) can be generalized to accommodate time-dependent volatilities and interest rates in analogy to (\ref{eq11a}) and (\ref{eq12a}), to yield
\begin{widetext}
\begin{equation}
\tilde{C}(S,T)=S\exp\bigg({1\over2}\int_0^T \gamma^2(\tau)s^2(\tau)-k(\tau)\,d\tau\bigg)N(\tilde{d}_1)-K\exp\bigg(-\int_0^T r(\tau)\,d\tau\bigg)N(\tilde{d}_2)\, ,
\label{eq22}
\end{equation}
with
\begin{align}
\tilde{d}_1&={\ln(S/K)+\int_0^T r(\tau)\,d\tau+\int_0^T\gamma^2(\tau)s^2(\tau)-{1\over2}k(\tau)\,d\tau\over\sqrt{\int_0^T\gamma^2(\tau)s^2(\tau)\,d\tau}}\, ,\\
\tilde{d}_2&={\ln(S/K)+\int_0^T r(\tau)\,d\tau-{1\over2}\int_0^T k(\tau)\,d\tau\over\sqrt{\int_0^T\gamma^2(\tau)s^2(\tau)\,d\tau}}\, .
\label{eq23}
\end{align}
\end{widetext}

%%%%%%%%%%%%%%%%%%%%%%%%%%%%%%%%%%%%%%%%%%%%%%%%%

\begin{figure*}[t!]
\centering
\includegraphics[width=0.8\textwidth]{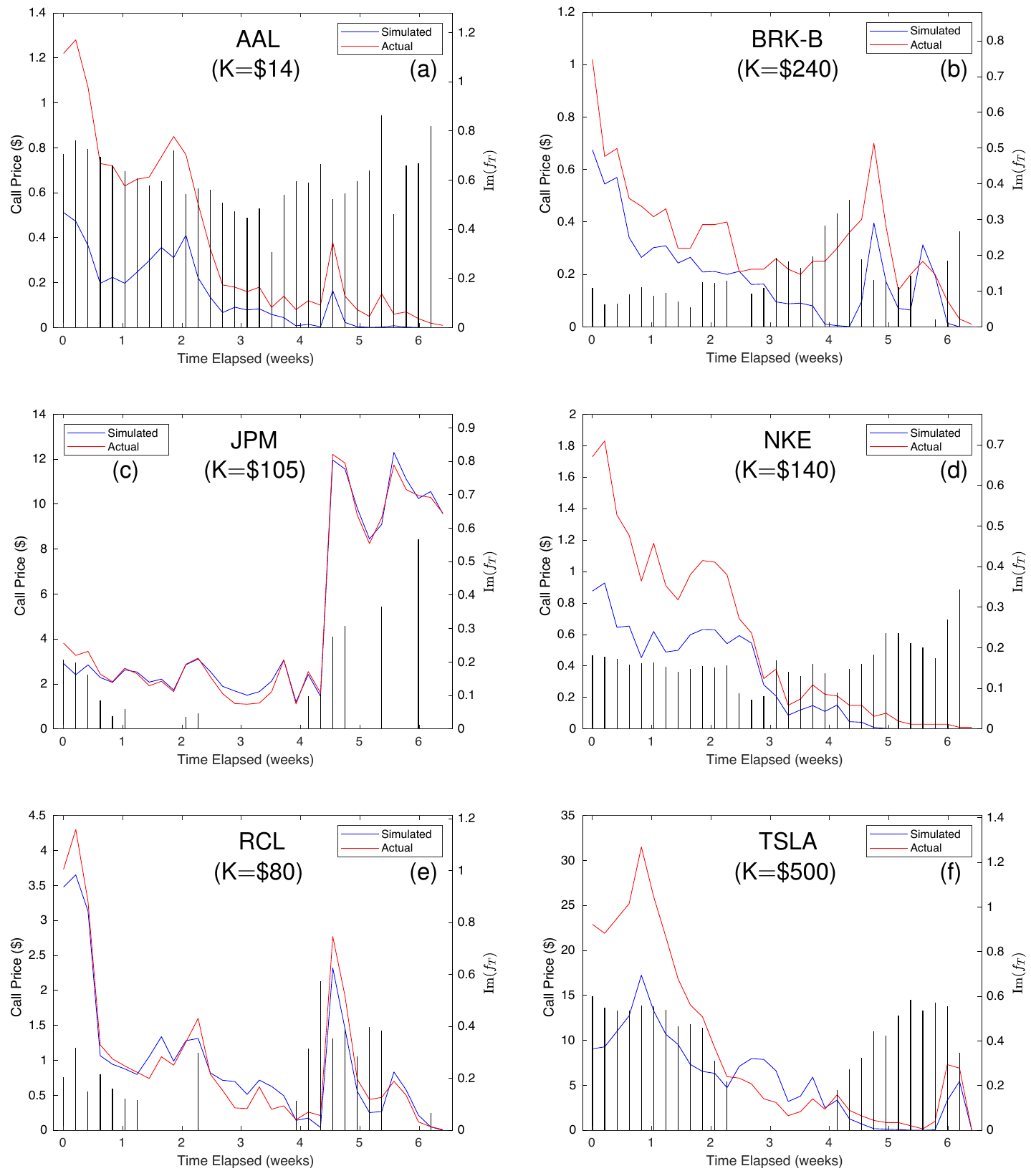}
\caption{Comparison between actual prices of European call options (shown in red) and prices calculated from the pricing formulae shown in Fig.~\ref{fig1} (blue). The vertical bars represent the values of $\mbox{Im}(f(T))$.  Gaps in these bars indicate that $\mbox{Im}(f(T))$ is itself imaginary, i.e.~that the optimized value of $f(T)$ is real.}
\label{fig3}
\end{figure*}

%%%%%%%%%%%%%%%%%%%%%%%%%%%%%%%%%%%%%%%%%%%%%%%%%

The results (\ref{eq17}) and (\ref{eq18}) obtained by Segal and Segal \cite{segal98} are based on mathematical development of non-commuting simultaneous processes, each governed by a Wiener process in the stochastic differential equation for the stock price (\ref{eq14a}).  We present here a test of these ideas against market data by determining the values of $\mbox{Im}(f(T))$, representing the additional part of the volatility for which the calculated and market prices in Fig.~\ref{fig1} coincide.  We have set $\gamma=1$, since an increasing $\gamma$ was found to increase the volatility of the option price towards either extremum in the time interval, i.e.~towards the initial time or the time of expiry. 

Results are shown in Fig.~\ref{fig3} for the stock options in Fig.~\ref{fig1}.  As is immediately evident, the value of $\mbox{Im}(f(T))$ increases with the difference between the actual and calculated option prices, and its magnitude is of order $\mbox{Re}(f(T))=\sigma$.  It is also interesting to note that $\mbox{Im}(f(T))$ is particularly sensitive to discrepancies at times closer to expiry, i.e.~at times when the expectation of the classical simulation converging to the actual price is greater. In other words, the divergence of option prices in the securities market obtained from the Black--Scholes model is compensated by the inclusion of a pseudo-Wiener process incorporating information that is not simultaneously observable with public information. This additional information may take the form of speculation or the reaction of large market participants to one or more events. The evidence we have presented points directly to the presence of inefficiencies in market pricing, which lends further credibility to the proposition that the efficient market hypothesis does not offer a complete description of the stock market at the level of individual securities or over short time scales, and that such inefficiencies have the potential to be exploited to make greater returns.

There is a noteworthy issue that results from an analysis of the under-performance of the option. Such instances yield entirely imaginary solutions for the value of $\mbox{Im}(f(T))$ and, hence, cannot be plotted on the same axis with instances that yield real solutions. It was found that the moduli of these imaginary solutions were, in general, much smaller than those of the real solutions. From a physical point of view, however, an imaginary value of $b$ in the coefficient of the generalized quantum Wiener process,  $\Phi((a + ib)c_t) =  \Phi(f(t))$, would be indistinguishable from a purely real contribution $a$ of the appropriate smaller magnitude, with $b = 0$. Since the deviation from the classical pricing model is greater when the quantum process has a non-vanishing contribution, this leads us to the conclusion that, while the probabilities of over-performance and under-performance may be equal at any given instant, the amount by which an option over-performs is greater than that by which it under-performs.If an option over-performs at a given instant in time, the discrepancy with respect to the classical model is greater than if it had under-performed at that instant, since $s=|f(T)|$ is naturally greater in the former case than in the latter.

\section{Summary and Conclusions}
\label{sec5}

We began by calculating option prices based on stocks from several sectors of the S\&P 500 from the classical Black--Scholes formulae with time-dependent coefficients.  By comparing these calculations with actual prices, we observed several trends that suggested the possibility of additional factors that determine the price of the option, but which cannot be observed (or actioned) concurrently with the publicly available information used in the classical theory.  Following the suggestion of Segal and Segal, the Black--Scholes model was expanded by using a free boson field with infinite degrees of freedom to represent this additional information. This statement was shown to take its usual meaning in the quantum sense, in the form of a non-zero commutation relation between the mappings denoting the two Wiener processes. 

The motivation behind using the underlying canonical commutation relations (in the Weyl form) was outlined by showing that the essential general mathematical structure for both systems is the same. The price of a call in the generalized development was determined, which yielded a generalized complex-valued volatility. These formulae were used to price the aforementioned options, with the imaginary part of the volatility shown to compensate for the difference between the market and classical Black--Scholes prices. This provides direct proof of a non-commuting correction to the classical model, providing market evidence that, over short enough time intervals, a securities market reflects essential quantum features, just like any other physical system.

\bibliographystyle{apsrev4-2}

\bibliography{cas-refs}

\end{document}

\title{Supplementary material for:\\ Quantum effects in an expanded Black--Scholes model}

\author{Anantya Bhatnagar}

\author{Dimitri D. Vvedensky}
 
\affiliation{The Blackett Laboratory, Imperial College London, London SW7 2AZ, United Kingdom}

\date{\today}

\begin{abstract}
This document presents additional comparisons between actual European call option prices with calculated prices using the original Black--Scholes model and the expanded model suggested by Segal and Segal.  Calculations have been carried out for stock options of 20 companies in different sectors of the S\&P 500, each at three strike prices.  These comparisons support the conclusions in the main manuscript by showing a wider range of price profiles and differences between the actual and calculated option prices.
\end{abstract}

\maketitle

\section{Introduction}

In the accompanying manuscript, comparisons are made between actual prices of European call options of stocks of several companies listed on the S\&P 500 and prices calculated from the classical Black--Scholes model.  These comparisons motivate an evaluation of the model proposed by Segal and Segal \cite{segal98}, who suggested the existence of additional processes that are not observable simultaneously with the processes governed by the Brownian motion in the Black--Scholes formulation.  The pricing formulae obtained by Segal and Segal, in which the volatility is complex, were used to determine the imaginary part of the volatility, and thus to indicate the influence of the additional processes.  In this document we provide additional comparisons between actual prices of European call options based on stocks of 20 S\&P 500 companies and the original Black--Scholes model.  The model of Segal and Segal is then used to account for the discrepancies between the market and Black--Scholes prices.

\section{Pricing Formulae}

The Black--Scholes pricing formulae \cite{black73,merton73,wilmott06,hull15} for a European call option are specified by five variables:

The strike price $K$

The current stock price $S$

The time $T$ to expiration

The risk-free interest rate $r$

The volatility $\sigma$

\noindent
The strike price and the time to expiration are set by the writer of the option.  The current stock price is available from several sources to whatever level of resolution desired.  We have used daily returns.

For the interest rate, we take the continuously compounded yield on a 3-month Treasury bill (T-bill) whose maturity date is closest to the expiry date of the option. T-bills are guaranteed by the government of the United States and are, therefore, considered to be free of default risk. Because the interest rates were so low, and the changes were essentially negligible over the 6-week period of our options window, we took the constant interest rate of 0.08\% for all our calculations.

We determine the percentage variance $\sigma$ for each stock from
\begin{equation}
\sigma=\beta{\mbox{VIX}\over100}\, .
\label{eq1}
\end{equation}
where VIX is the ticker symbol for the volatility index of the Chicago Board Options Exchange, which is a real-time measure of the volatility based on S\&P 500 index options with near-term expiration dates.  The factor $\beta$ is a measure of the volatility of a stock compared to the volatility of all other stocks in a particular index, in our case, the S\&P 500 index.  The values of $\beta$ \cite{beta} used for the calculations reported here are compiled in Table~\ref{table1}.

\begin{table}[b]
\renewcommand{\tabcolsep}{0.2cm}
\caption{Values of $\beta$ used in (\ref{eq1}) to calculate the volatility for the European call options in based on the stocks of the 20 companies from the S\&P 500 whose prices are calculated from (\ref{eq2}) and (\ref{eq3}).}
\vskip12pt
\centering
\resizebox{\textwidth}{!}{\begin{tabular}{| l l l || l l l |} 
\hline\hline
Ticker & \multicolumn{1}{c}{Company Name} & \multicolumn{1}{c||}{$\beta$} & Ticker & \multicolumn{1}{c}{Company Name} &  \multicolumn{1}{c|}{$\beta$} \\
\hline
\hline
AAL & American Airlines Group & 1.71 & AAPL & Apple Inc. & 1.36 \\
AMD & Advanced Micro Devices, Inc. & 2.32 & AMZN &Amazon.com, Inc. & 1.31 \\
BA & Boeing Co. & 1.41 & BAC & Bank of America Corp. & 1.57 \\
BRK-B & Berkshire Hathaway Inc. Class B & 0.84 & C & Citigroup Inc. & 1.82 \\
GS & Goldman Sachs Group Inc, & 1.42 & INTC & Intel Corporation & 0.68 \\
JPM & JP Morgan Chase \& Co. & 1.12 & M & Macy's Inc. & 1.82 \\
MAR & Marriott International Inc. & 1.68 & NFLX & Netflix Inc. & 0.98 \\
NKE & Nike Inc. & 0.82 & PFE & Pfizer Inc. & 0.72 \\
RCL & Royal Caribbean Cruises Ltd & 2.76 & TSLA & Tesla Inc. & 1.97\\
WMT & Walmart Inc. & 0.40 & ZM & Zoom Video Communications Inc. & 1.05 \\
\hline
\end{tabular}}
\label{table1}
\end{table}

For a constant risk-free interest rate and a time-dependent volatility, the Black--Scholes pricing formulae are \cite{wilmott06}:
\begin{equation}
C^{(1)}(S,T)=SN(d_1^{(1)})-Ke^{-rT}N(d_2^{(1)})\, ,
\label{eq2}
\end{equation}
where the new functions $d_1^{(1)}$ and $d_2^{(1)}$ are now given by
\begin{gather}
\begin{aligned}
d_1^{(1)}&={\ln(S/K)+rT+{1\over2}\int_0^T\sigma^2(\tau)\,d\tau\over\sqrt{\int_0^T\sigma^2(\tau)\,d\tau}}\, ,\\
\noalign{\vskip6pt}
d_2^{(1)}&={\ln(S/K)+rT-{1\over2}\int_0^T\sigma^2(\tau)\,d\tau\over\sqrt{\int_0^T\sigma^2(\tau)\,d\tau}}\, .
\end{aligned}
\label{eq3}
\end{gather}

The corresponding pricing formulas obtained by Segal and Segal (modified for a time-dependent volatility) are \cite{segal98}:
\begin{equation}
C^{(3)}(S,T)=S\exp\bigg\{{1\over2}\int_0^T \big[\gamma^2(\tau)s^2(\tau)-k(\tau)\big]\,d\tau\bigg\}N(d^{(3)}_1)-Ke^{-rT}N(d^{(3)}_2)\, ,
\label{eq4}
\end{equation}
with
\begin{equation}
\begin{gathered}
d^{(3)}_1={\ln(S/K)+rT+\int_0^T\big[\gamma^2(\tau)s^2(\tau)-{1\over2}k(\tau)\big]\,d\tau\over\sqrt{\int_0^T\gamma^2(\tau)s^2(\tau)\,d\tau}}\, ,\\
\noalign{\vskip6pt}
d^{(3)}_2={\ln(S/K)+rT-{1\over2}\int_0^T k(\tau)\,d\tau\over\sqrt{\int_0^T\gamma^2(\tau)s^2(\tau)\,d\tau}}\, ,
\end{gathered}
\label{eq5}
\end{equation}
where the real part of $f(t)$ can be regarded as the process representing public information, whereas the imaginary part corresponds to a process that cannot be observed simultaneously with public information.  Here, $s=|f(T)|$, $\gamma\ge1$, and $k=s^2$ or $k=0$ for Wiener or serial correlated processes, respectively.

\section{\hskip-9pt Comparisons between Actual and Calculated Option Prices}

The main result in the accompanying manuscript is that the imaginary part of the volatility in (\ref{eq4}) and (\ref{eq5}) can alleviate the differences between actual option prices and calculations based on (\ref{eq2}) and (\ref{eq3}) where the calculations underestimate the actual price.  Our comparisons are based on options on 20 stocks from different sectors of the S\&P 500.  There are 11 sectors in the S\&P 500. As of December 31, 2020 \cite{sectors}, the order of the sectors based on size, which is shown in parentheses, is as follows:

Information technology (27.6\%)

Health care (13.5\%)

Consumer discretionary (12.7\%)

Communication services (10.8\%)

Financials (10.4\%)

Industrials (8.4\%)

Consumer staples (6.5\%)

Utilities (2.8\%)

Materials (2.6\%)

Real estate (2.4\%)

Energy (2.3\%).

\begin{table}[b!]
\renewcommand{\tabcolsep}{0.2cm}
\caption{The sectors within the S\&P 500 of the 20 stocks whose call option prices have been calculated.}
\centering
\begin{tabular}{| l l |} 
\hline\hline
\multicolumn{1}{|c}{Company Name} & \multicolumn{1}{c||}{Sector} \\
\hline
\hline
American Airlines Group & Industrials \\
Apple Inc. & Information Technology \\
Advanced Micro Devices Inc. & Information Technology \\
Amazon.com Inc. & Consumer Discretionary \\
Boeing Co. & Industrials \\
Bank of America Corp. & Financials \\
Berkshire Hathaway Inc.~Class B & Financials \\
Citigroup Inc. & Financials \\
Goldman Sachs Group Inc. & Financials \\
Intel Corporation & Information Technology \\
JP Morgan Chase \& Co. & Financials \\
Macy's Inc. & Consumer Discretionary \\
Marriott International Inc. & Consumer Discretionary \\
Netflix Inc. & Communication Services \\
Nike Inc. & Consumer Discretionary \\
Pfizer Inc. & Health Care \\
Royal Caribbean Cruises Ltd & Consumer Discretionary \\
Tesla Inc. & Consumer Discretionary \\
Walmart Inc. & Consumer Staples \\
Zoom Video Communications Inc. & Information Technology \\
\hline
\end{tabular}
\label{table2}
\end{table}

The sectors of the 20 stocks whose option prices are calculated are shown on Table~\ref{table2}.  The following figures show the comparisons between the actual prices of European call options based on the stocks of the 20 S\&P 500 companies in Tables~\ref{table1} and \ref{table2} and the original Black--Scholes model based on (\ref{eq2}) and (\ref{eq3}), and between the expanded Black--Scholes model based on (\ref{eq4}) and (\ref{eq5}).  In each figure, historical data \cite{bloomberg} was used for the closing price of the call option on each trading day over the 6-week period from October 8, 2020 to November 20, 2020.  Three strike prices are shown for each option.

\begin{figure*}[p!] 
\includegraphics[width=\textwidth]{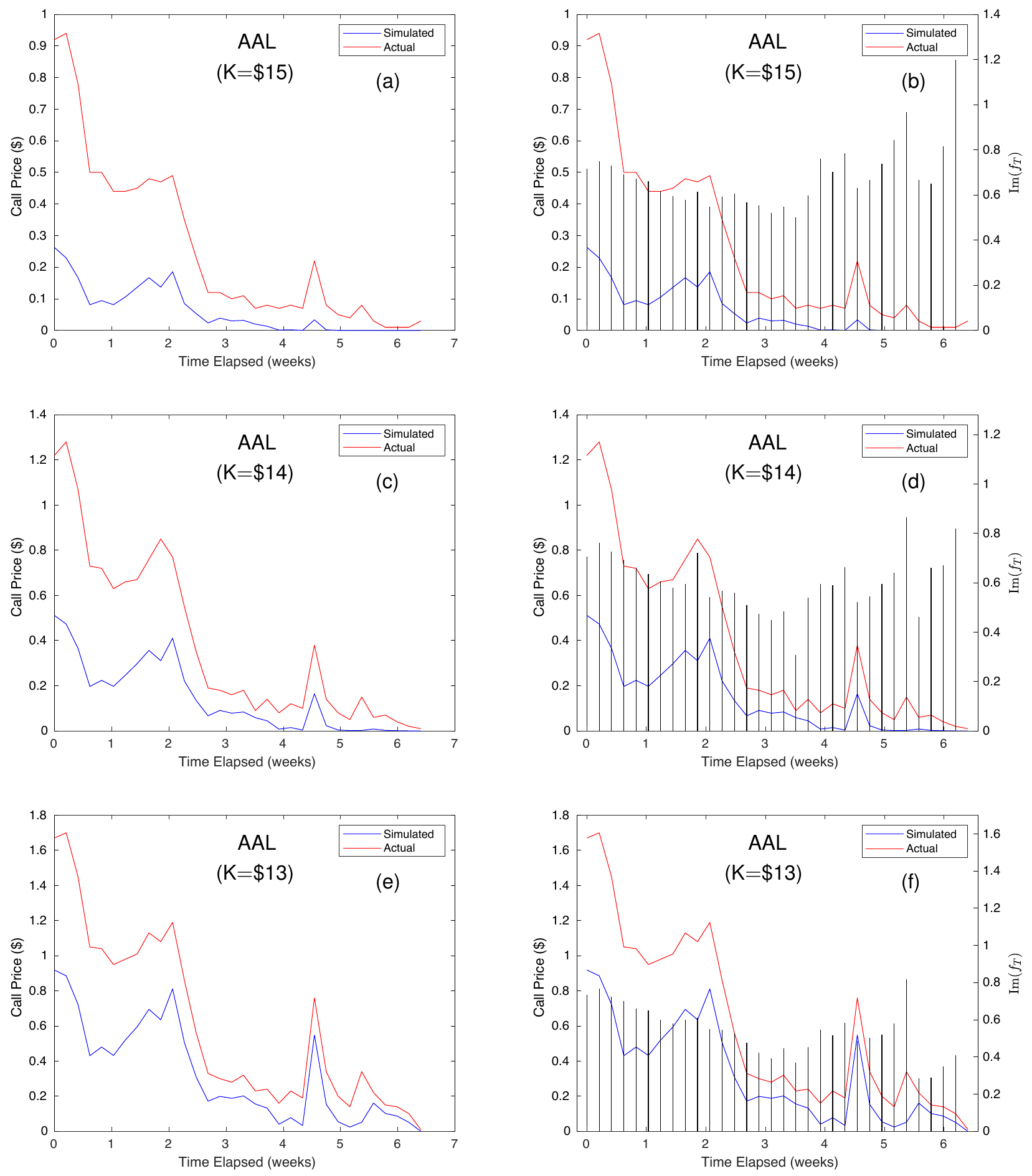}
\caption{Comparison between actual prices of a European call option (red lines) for stocks of the American Airlines Group and (a,c,e) the original Black--Scholes model based on the pricing formulae (\ref{eq2}) and (\ref{eq3}) (blue lines), and (b,c,d) the expanded Black--Scholes model based on the pricing formulae (\ref{eq4}) and (\ref{eq5}) of the expanded Black--Scholes model proposed by Segal and Segal (black vertical lines).}
\end{figure*}

\begin{figure*}[p!] 
\includegraphics[width=\textwidth]{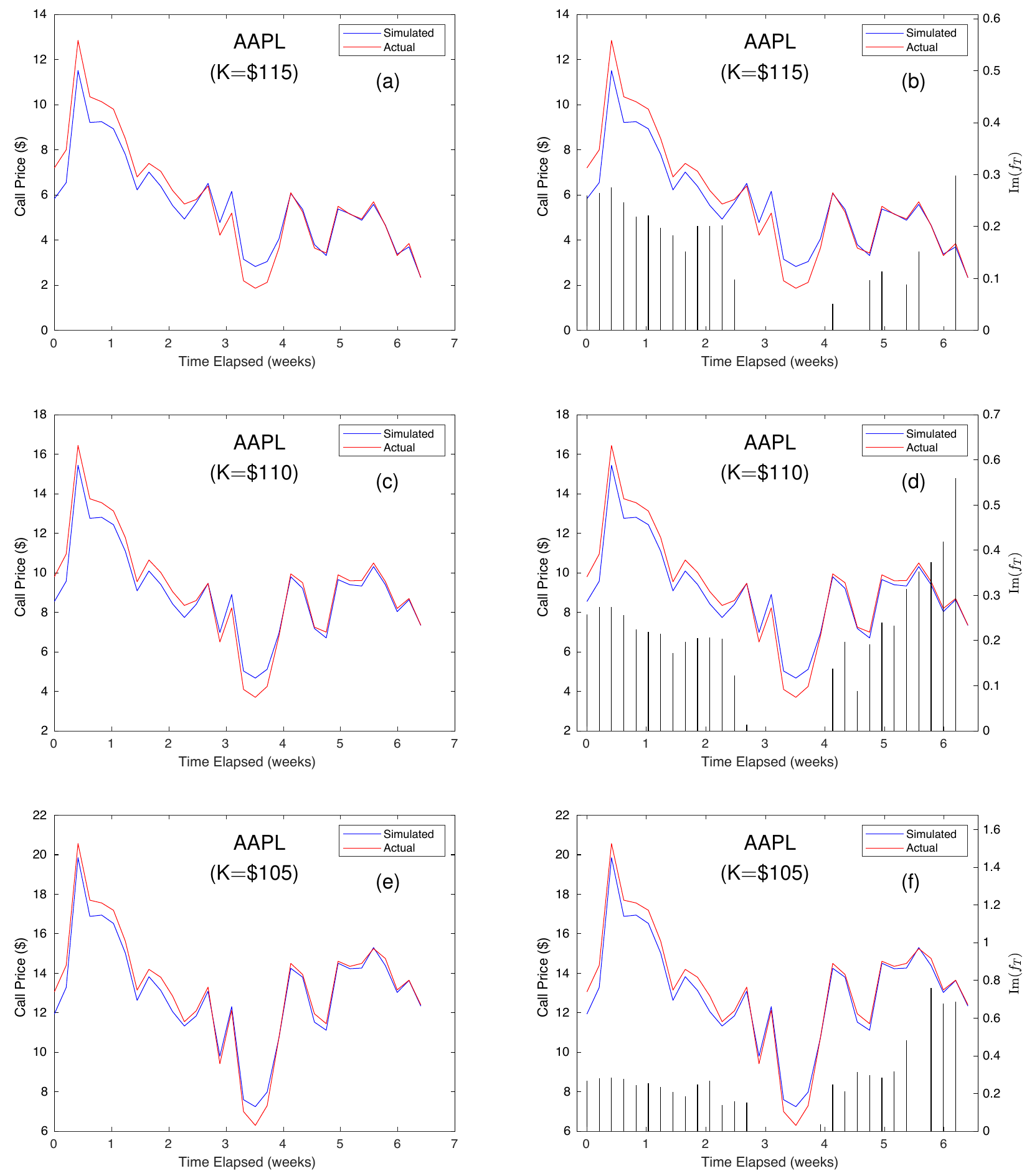}
\caption{Comparison between actual prices of a European call option (red lines) for stocks of Apple Inc.~ and (a,c,e) the original Black--Scholes model based on the pricing formulae (\ref{eq2}) and (\ref{eq3}) (blue lines), and (b,c,d) the expanded Black--Scholes model based on the pricing formulae (\ref{eq4}) and (\ref{eq5}) of the expanded Black--Scholes model proposed by Segal and Segal (black vertical lines).}
\end{figure*}

\begin{figure*}[p!] 
\includegraphics[width=\textwidth]{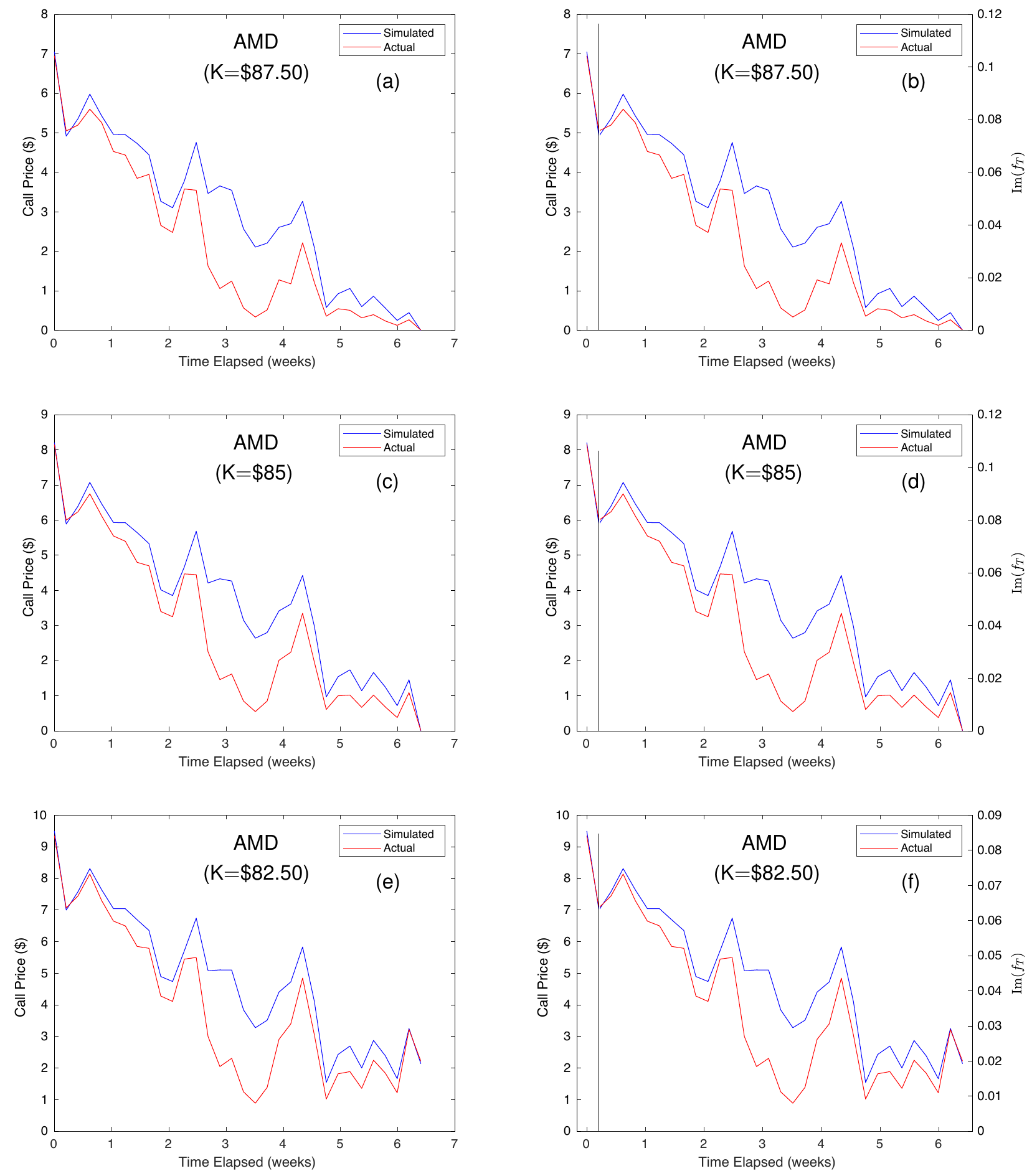}
\caption{Comparison between actual prices of a European call option (red lines) for stocks of Advanced Micro Devices Inc.~and (a,c,e) the original Black--Scholes model based on the pricing formulae (\ref{eq2}) and (\ref{eq3}) (blue lines), and (b,c,d) the expanded Black--Scholes model based on the pricing formulae (\ref{eq4}) and (\ref{eq5}) of the expanded Black--Scholes model proposed by Segal and Segal (black vertical lines).}
\end{figure*}

\begin{figure*}[p!] 
\includegraphics[width=\textwidth]{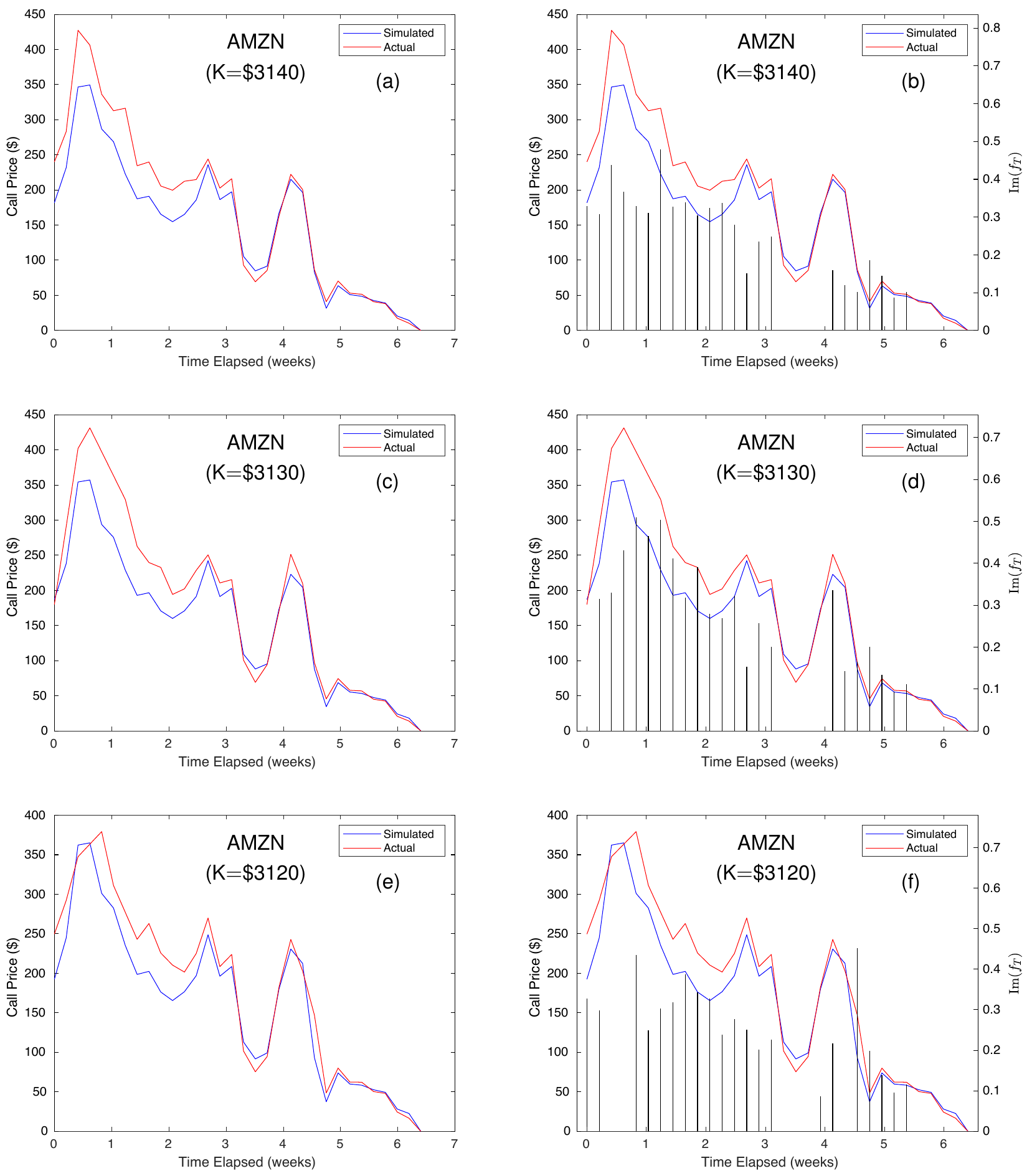}
\caption{Comparison between actual prices of a European call option (red lines) for stocks of Amazon.com Inc.~and (a,c,e) the original Black--Scholes model based on the pricing formulae (\ref{eq2}) and (\ref{eq3}) (blue lines), and (b,c,d) the expanded Black--Scholes model based on the pricing formulae (\ref{eq4}) and (\ref{eq5}) of the expanded Black--Scholes model proposed by Segal and Segal (black vertical lines).}
\end{figure*}

\begin{figure*}[p!] 
\includegraphics[width=\textwidth]{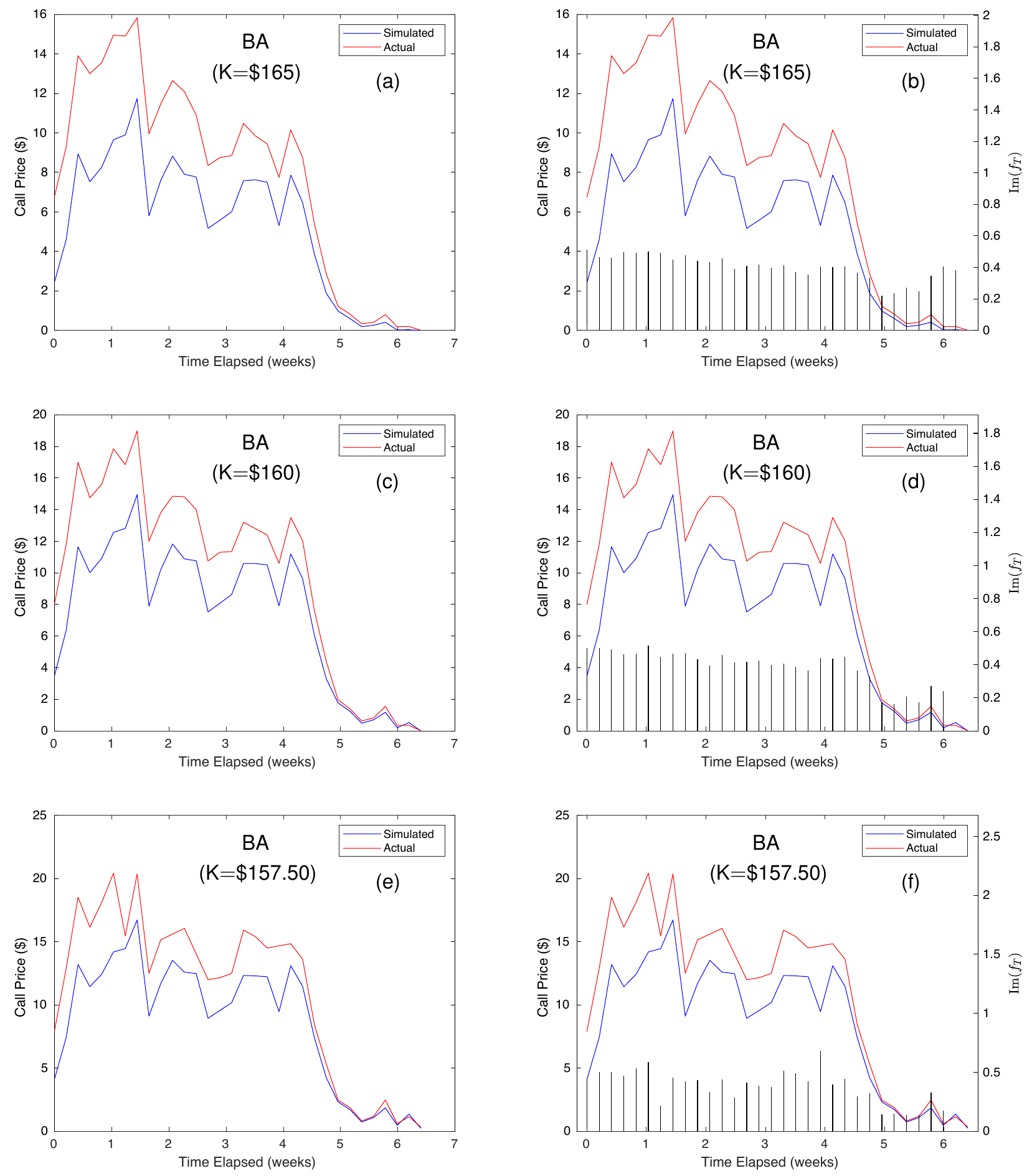}
\caption{Comparison between actual prices of a European call option (red lines) for stocks of the Boeing Co.~and (a,c,e) the original Black--Scholes model based on the pricing formulae (\ref{eq2}) and (\ref{eq3}) (blue lines), and (b,c,d) the expanded Black--Scholes model based on the pricing formulae (\ref{eq4}) and (\ref{eq5}) of the expanded Black--Scholes model proposed by Segal and Segal (black vertical lines).}
\end{figure*}

\begin{figure*}[p!] 
\includegraphics[width=\textwidth]{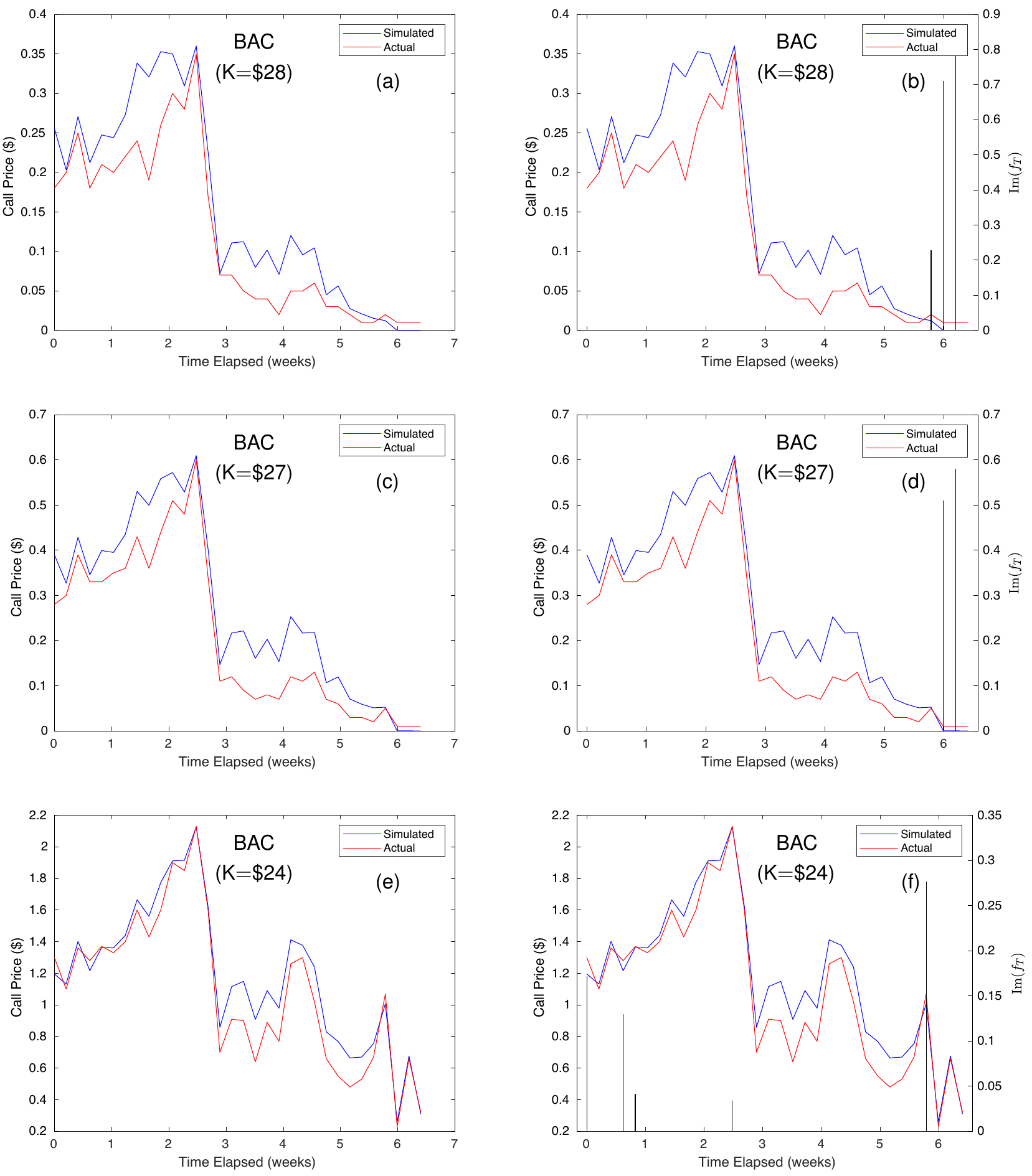}
\caption{Comparison between actual prices of a European call option (red lines) for stocks of the Bank of America Corp.~and (a,c,e) the original Black--Scholes model based on the pricing formulae (\ref{eq2}) and (\ref{eq3}) (blue lines), and (b,c,d) the expanded Black--Scholes model based on the pricing formulae (\ref{eq4}) and (\ref{eq5}) of the expanded Black--Scholes model proposed by Segal and Segal (black vertical lines).}
\end{figure*}

\begin{figure*}[p!] 
\includegraphics[width=\textwidth]{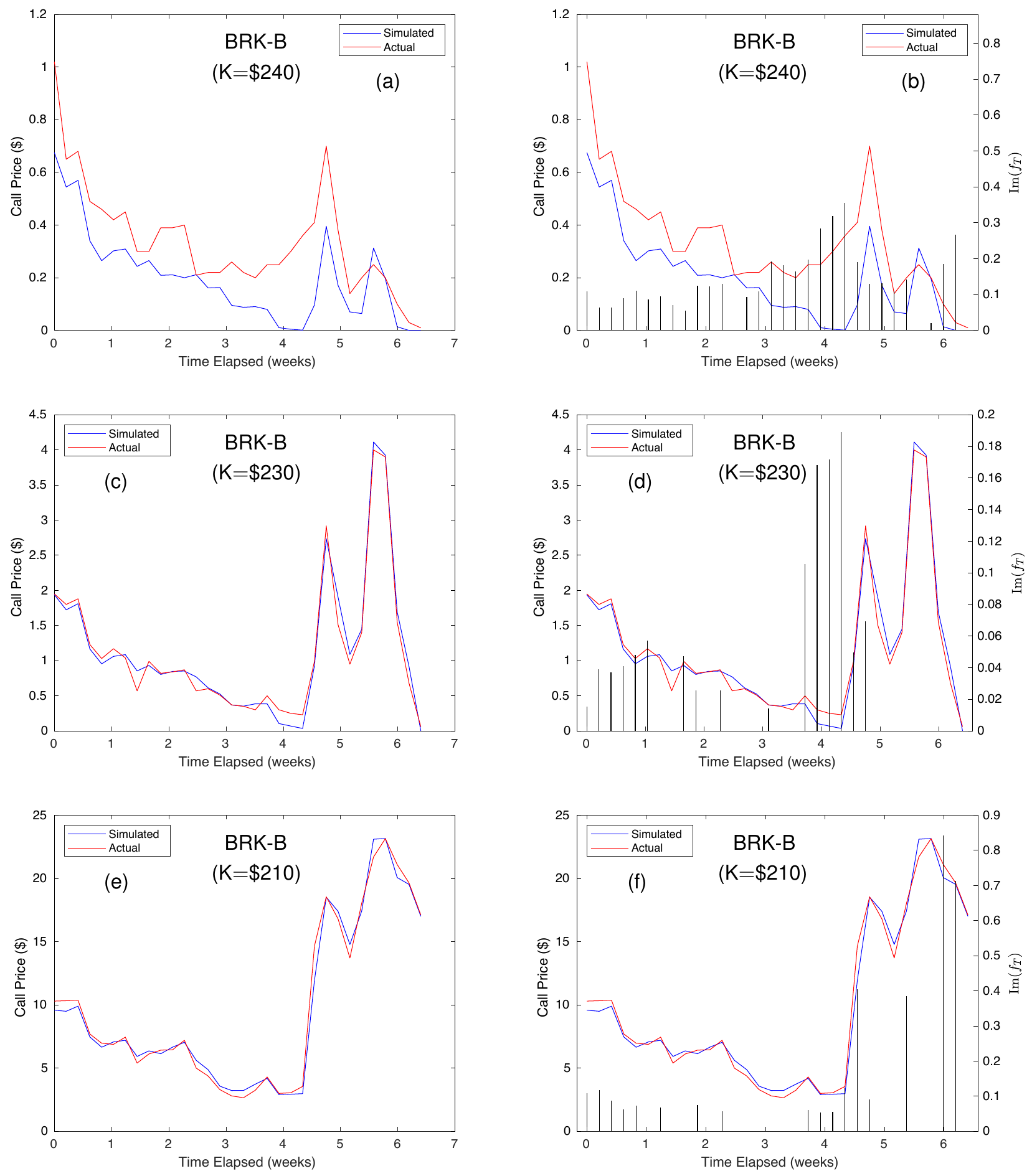}
\caption{Comparison between actual prices of a European call option (red lines) for stocks of Berkshire Hathaway Inc.~Class B and (a,c,e) the original Black--Scholes model based on the pricing formulae (\ref{eq2}) and (\ref{eq3}) (blue lines), and (b,c,d) the expanded Black--Scholes model based on the pricing formulae (\ref{eq4}) and (\ref{eq5}) of the expanded Black--Scholes model proposed by Segal and Segal (black vertical lines).}
\end{figure*}

\begin{figure*}[p!] 
\includegraphics[width=\textwidth]{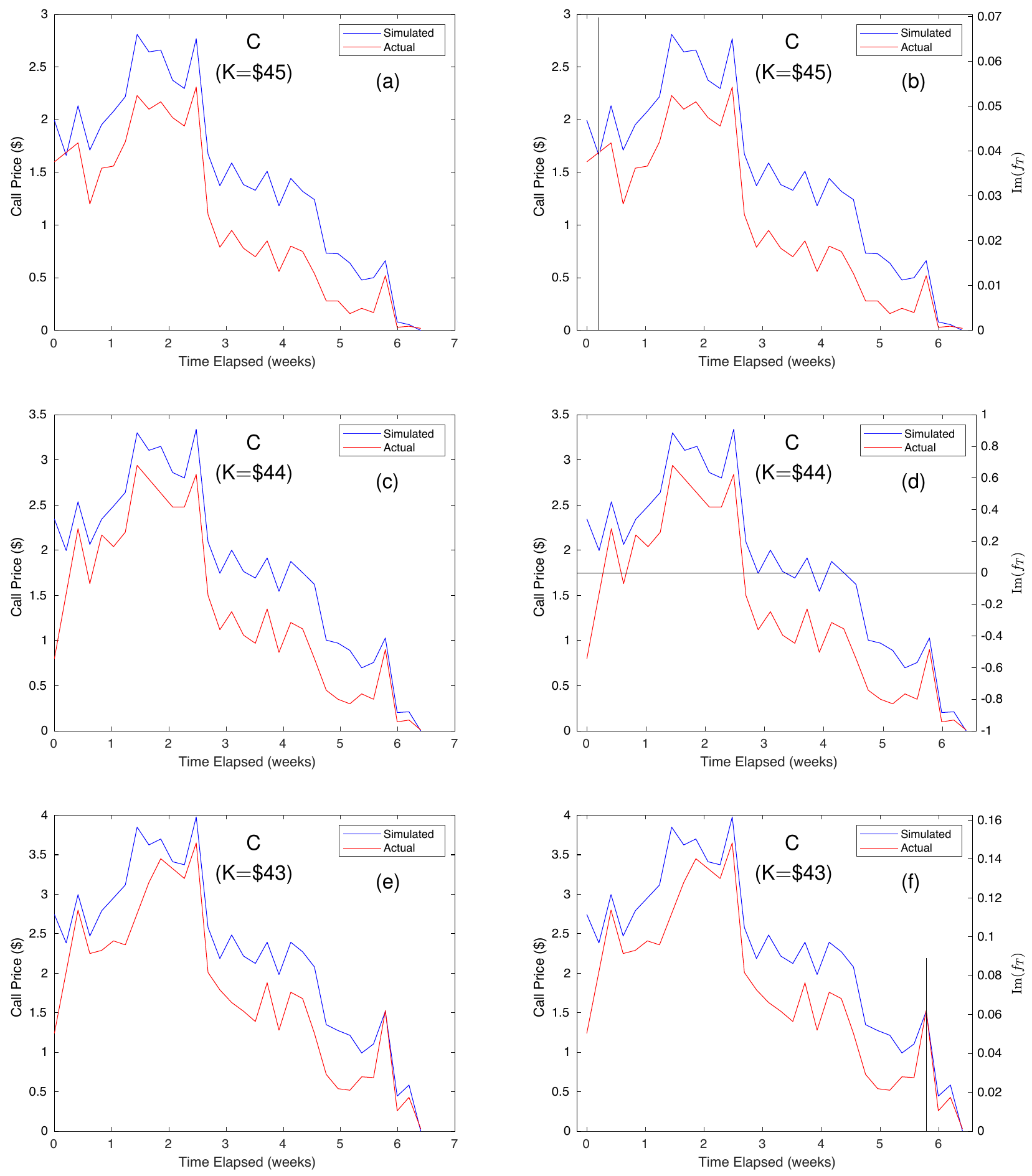}
\caption{Comparison between actual prices of a European call option (red lines) for stocks of Citigroup Inc.~Class B and (a,c,e) the original Black--Scholes model based on the pricing formulae (\ref{eq2}) and (\ref{eq3}) (blue lines), and (b,c,d) the expanded Black--Scholes model based on the pricing formulae (\ref{eq4}) and (\ref{eq5}) of the expanded Black--Scholes model proposed by Segal and Segal (black vertical lines).}
\end{figure*}

\begin{figure*}[p!] 
\includegraphics[width=\textwidth]{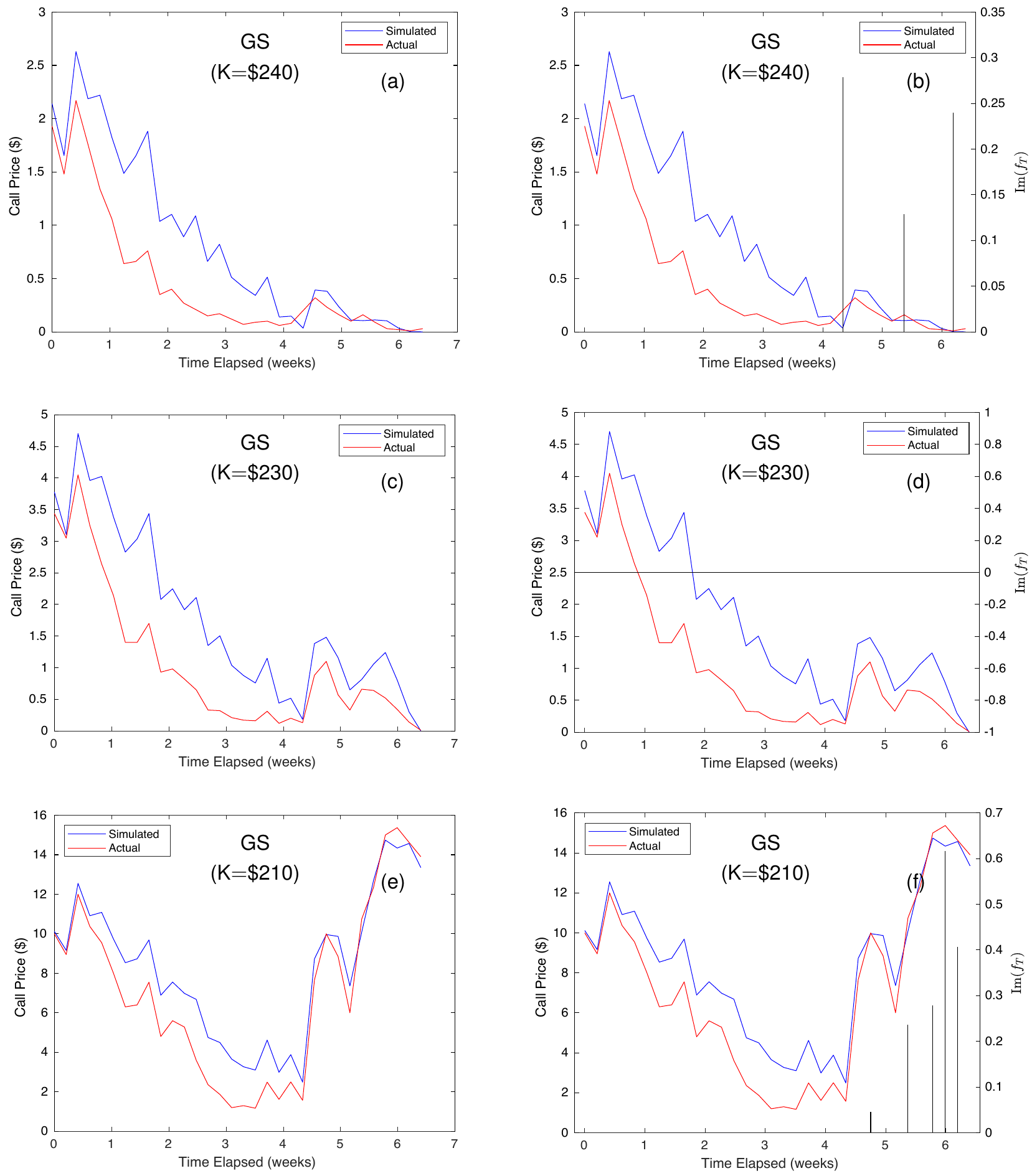}
\caption{Comparison between actual prices of a European call option (red lines) for stocks of the Goldman Sachs Group Inc.~and (a,c,e) the original Black--Scholes model based on the pricing formulae (\ref{eq2}) and (\ref{eq3}) (blue lines), and (b,c,d) the expanded Black--Scholes model based on the pricing formulae (\ref{eq4}) and (\ref{eq5}) of the expanded Black--Scholes model proposed by Segal and Segal (black vertical lines).}
\end{figure*}

\begin{figure*}[p!] 
\includegraphics[width=\textwidth]{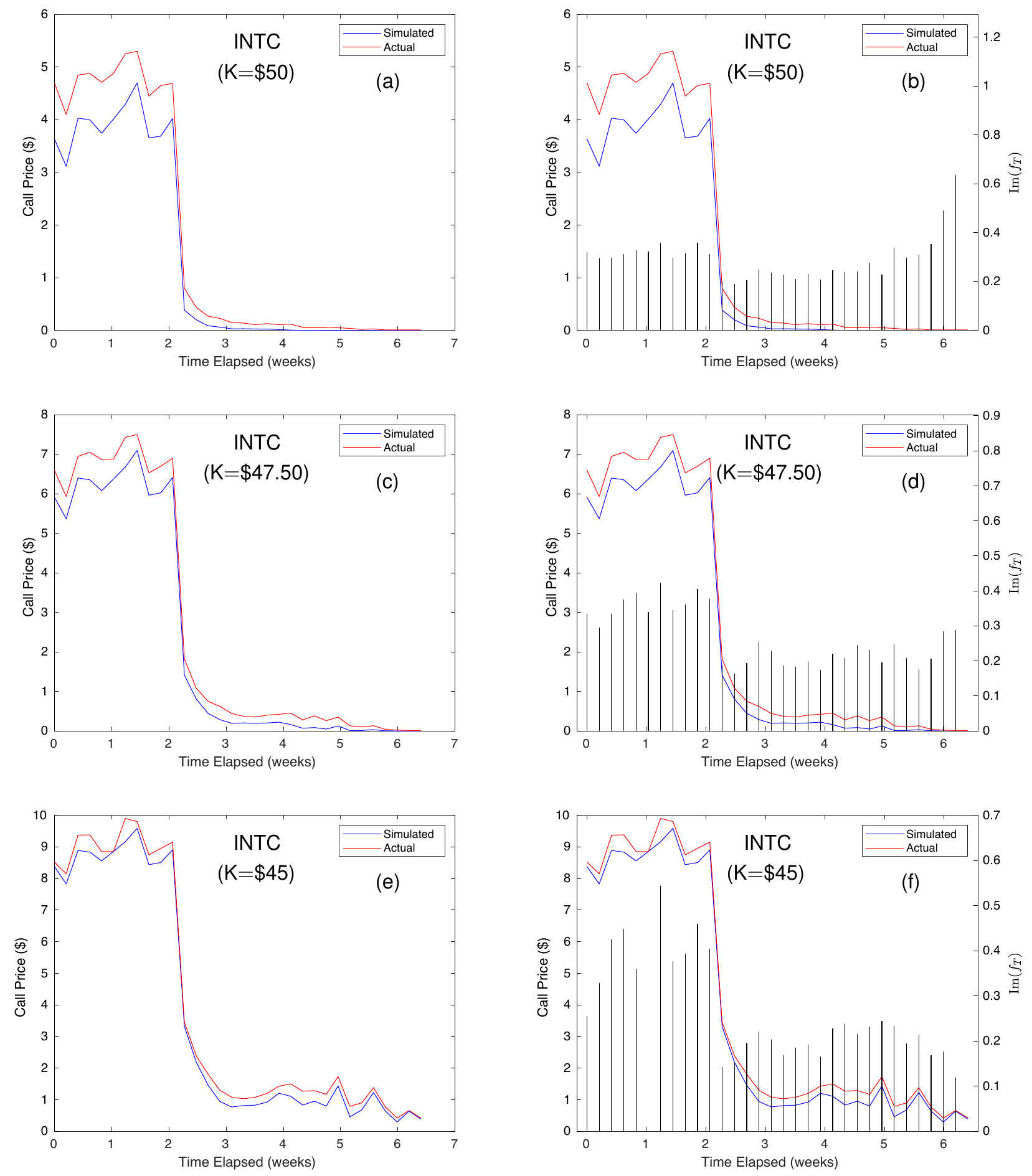}
\caption{Comparison between actual prices of a European call option (red lines) for stocks of the Intel Corporation and (a,c,e) the original Black--Scholes model based on the pricing formulae (\ref{eq2}) and (\ref{eq3}) (blue lines), and (b,c,d) the expanded Black--Scholes model based on the pricing formulae (\ref{eq4}) and (\ref{eq5}) of the expanded Black--Scholes model proposed by Segal and Segal (black vertical lines).}
\end{figure*}

\begin{figure*}[p!] 
\includegraphics[width=\textwidth]{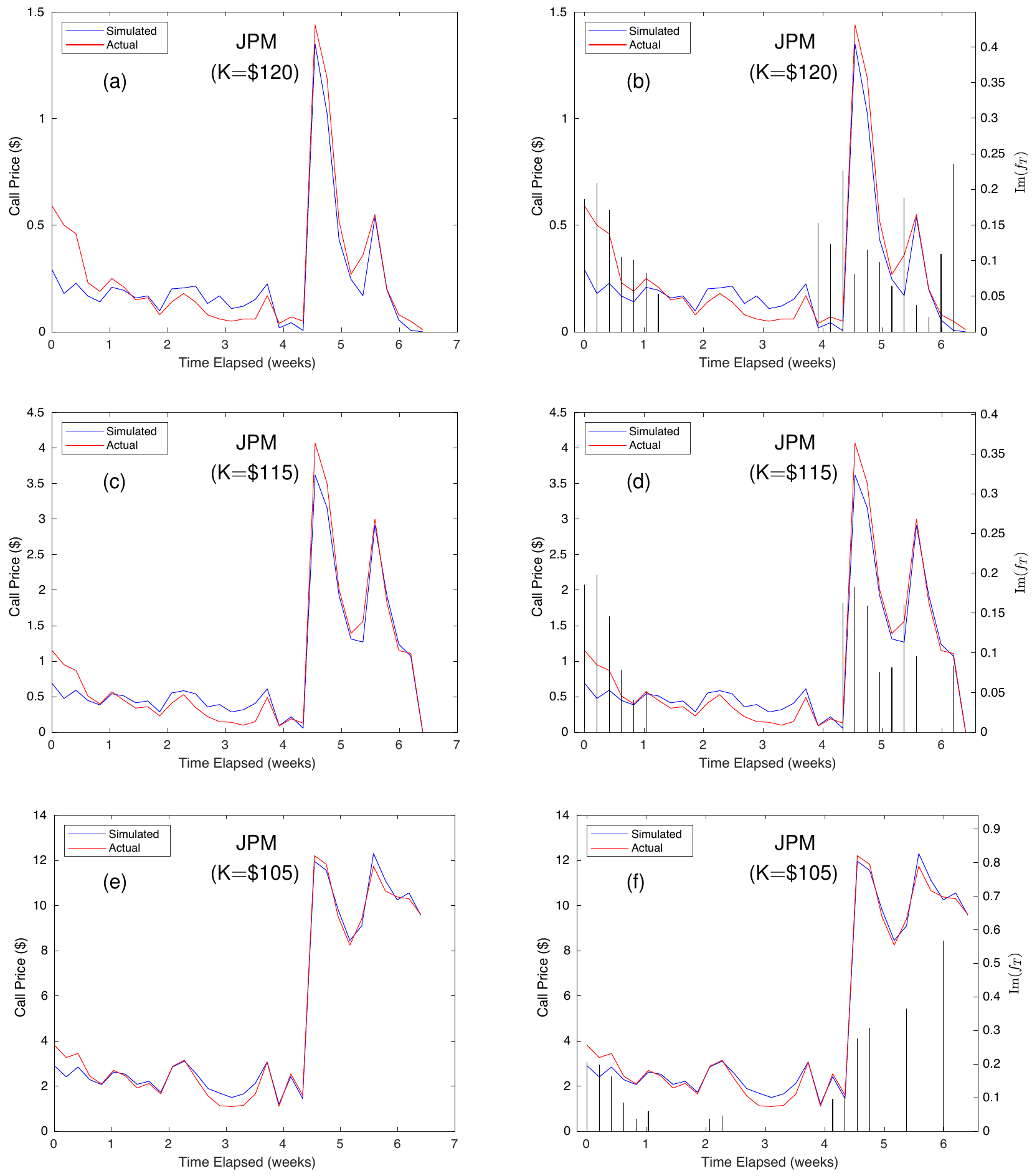}
\caption{Comparison between actual prices of a European call option (red lines) for stocks of JP Morgan Chase \& Co.~and (a,c,e) the original Black--Scholes model based on the pricing formulae (\ref{eq2}) and (\ref{eq3}) (blue lines), and (b,c,d) the expanded Black--Scholes model based on the pricing formulae (\ref{eq4}) and (\ref{eq5}) of the expanded Black--Scholes model proposed by Segal and Segal (black vertical lines).}
\end{figure*}

\begin{figure*}[p!] 
\includegraphics[width=\textwidth]{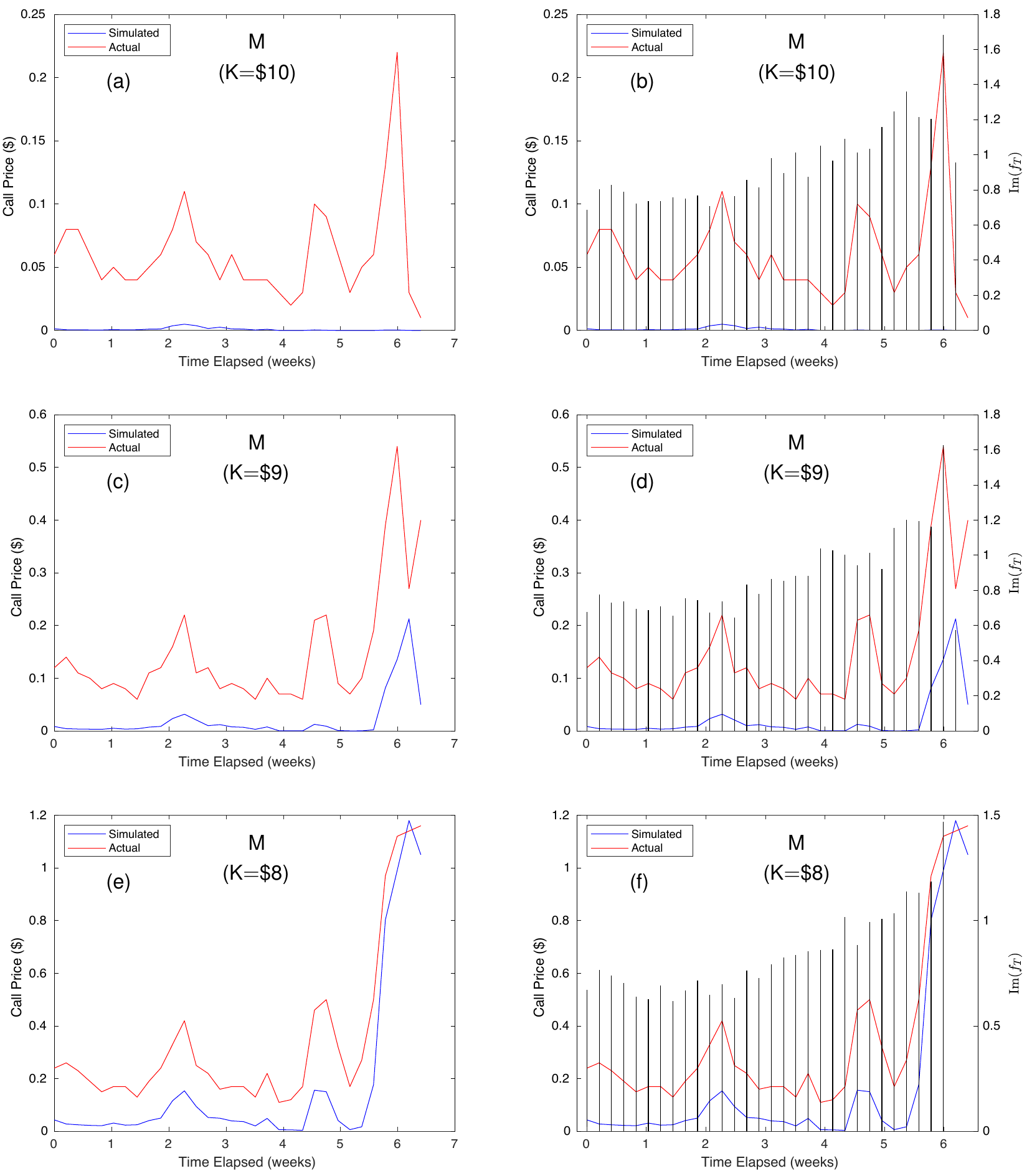}
\caption{Comparison between actual prices of a European call option (red lines) for stocks of Macy's Inc.~and (a,c,e) the original Black--Scholes model based on the pricing formulae (\ref{eq2}) and (\ref{eq3}) (blue lines), and (b,c,d) the expanded Black--Scholes model based on the pricing formulae (\ref{eq4}) and (\ref{eq5}) of the expanded Black--Scholes model proposed by Segal and Segal (black vertical lines).}
\end{figure*}

\begin{figure*}[p!] 
\includegraphics[width=\textwidth]{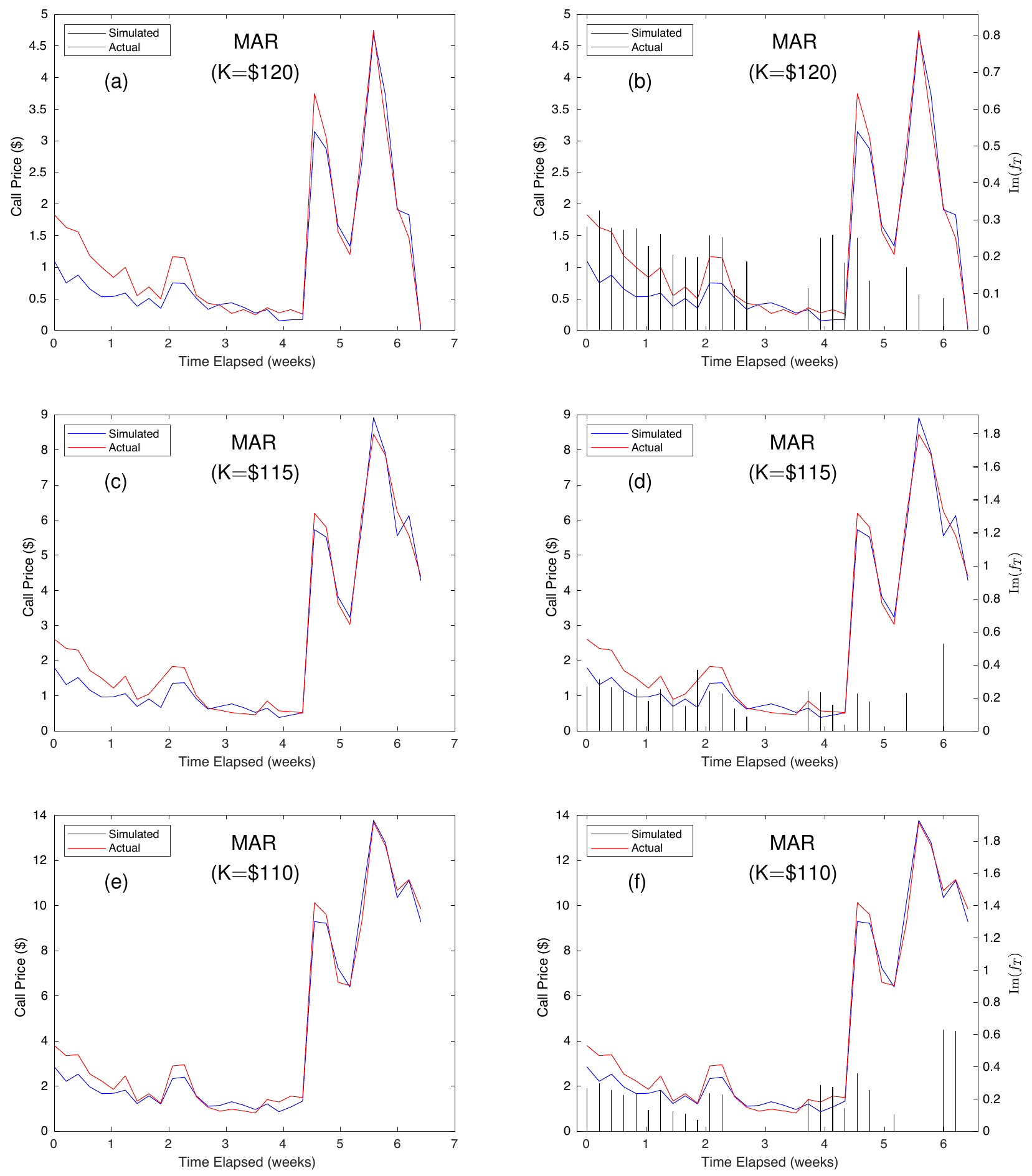}
\caption{Comparison between actual prices of a European call option (red lines) for stocks of Marriott International Inc.~and (a,c,e) the original Black--Scholes model based on the pricing formulae (\ref{eq2}) and (\ref{eq3}) (blue lines), and (b,c,d) the expanded Black--Scholes model based on the pricing formulae (\ref{eq4}) and (\ref{eq5}) of the expanded Black--Scholes model proposed by Segal and Segal (black vertical lines).}
\end{figure*}

\begin{figure*}[p!] 
\includegraphics[width=\textwidth]{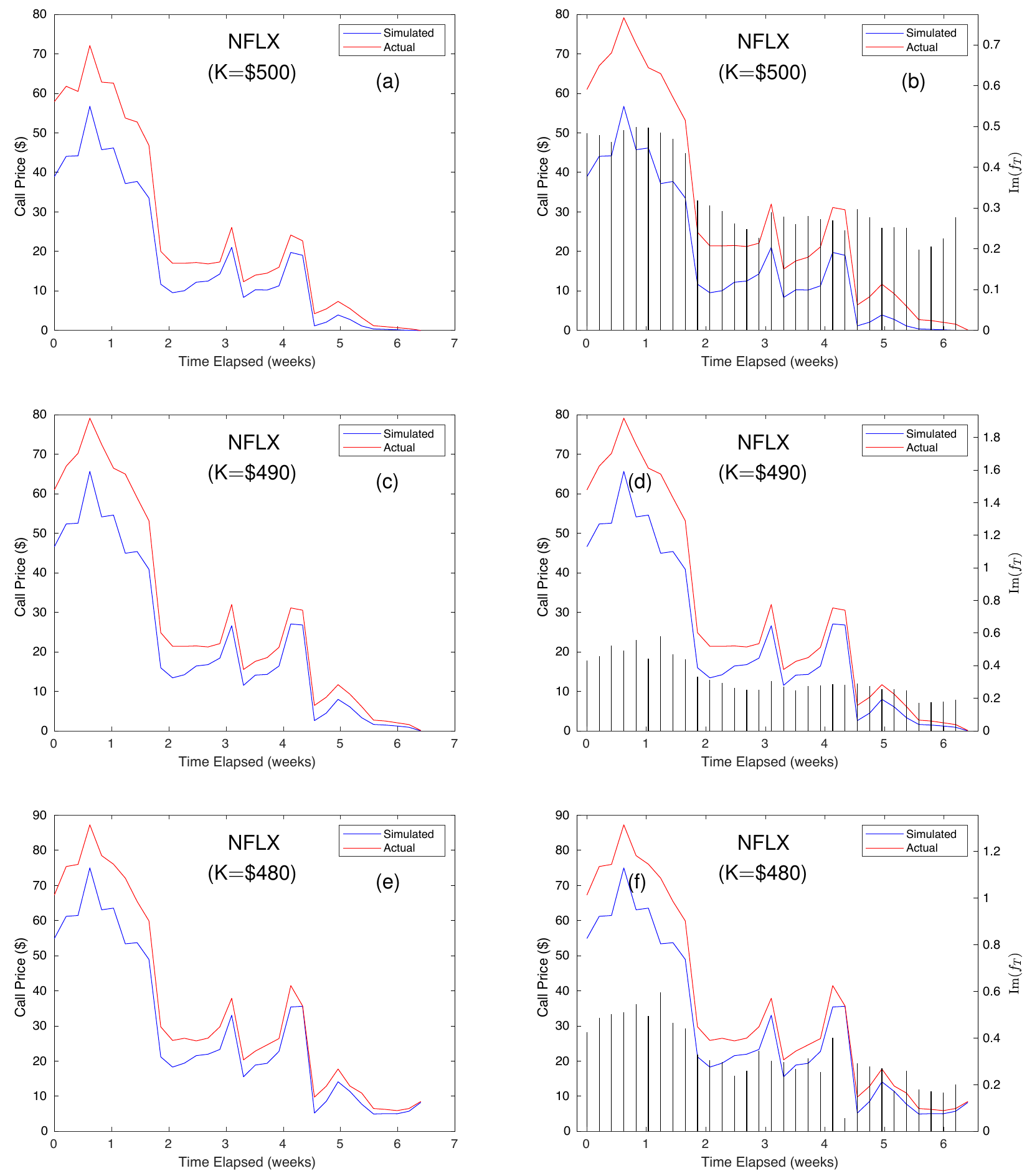}
\caption{Comparison between actual prices of a European call option (red lines) for stocks of Netflix Inc.~and (a,c,e) the original Black--Scholes model based on the pricing formulae (\ref{eq2}) and (\ref{eq3}) (blue lines), and (b,c,d) the expanded Black--Scholes model based on the pricing formulae (\ref{eq4}) and (\ref{eq5}) of the expanded Black--Scholes model proposed by Segal and Segal (black vertical lines).}
\end{figure*}

\begin{figure*}[p!] 
\includegraphics[width=\textwidth]{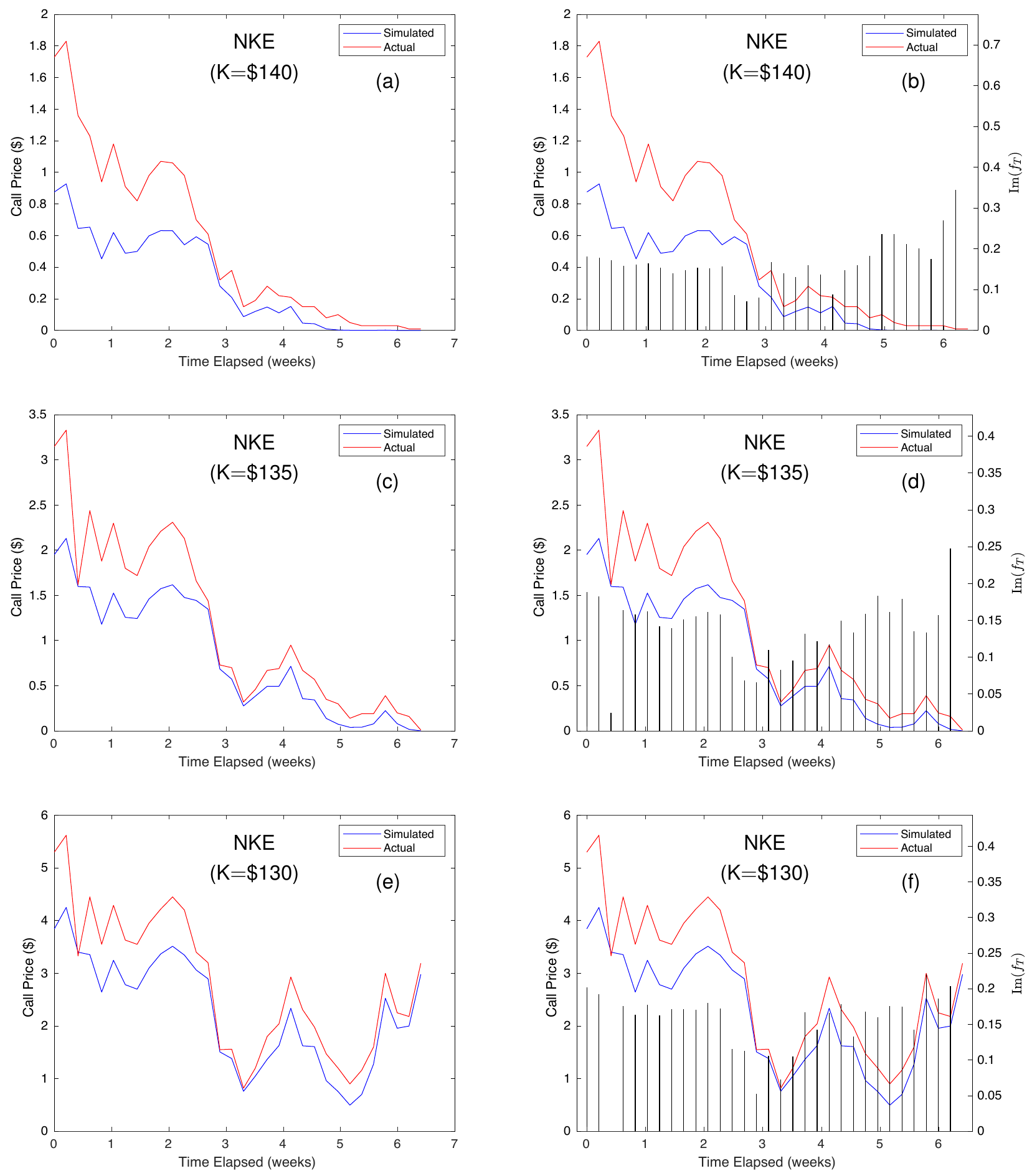}
\caption{Comparison between actual prices of a European call option (red lines) for stocks of Nike Inc.~and (a,c,e) the original Black--Scholes model based on the pricing formulae (\ref{eq2}) and (\ref{eq3}) (blue lines), and (b,c,d) the expanded Black--Scholes model based on the pricing formulae (\ref{eq4}) and (\ref{eq5}) of the expanded Black--Scholes model proposed by Segal and Segal (black vertical lines).}
\end{figure*}

\begin{figure*}[p!] 
\includegraphics[width=\textwidth]{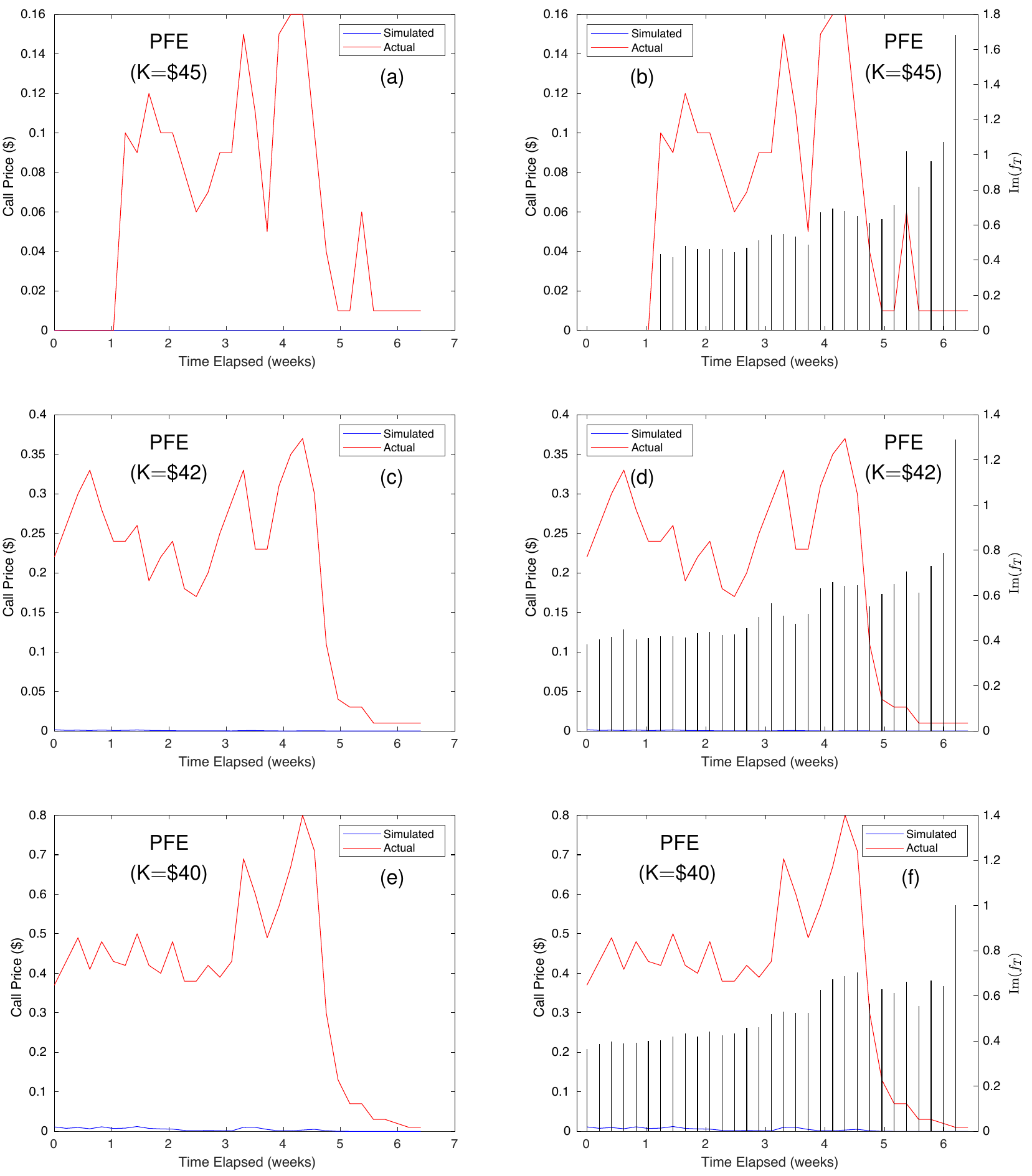}
\caption{Comparison between actual prices of a European call option (red lines) for stocks of Pfizer Inc.~and (a,c,e) the original Black--Scholes model based on the pricing formulae (\ref{eq2}) and (\ref{eq3}) (blue lines), and (b,c,d) the expanded Black--Scholes model based on the pricing formulae (\ref{eq4}) and (\ref{eq5}) of the expanded Black--Scholes model proposed by Segal and Segal (black vertical lines).}
\end{figure*}

\begin{figure*}[p!] 
\includegraphics[width=\textwidth]{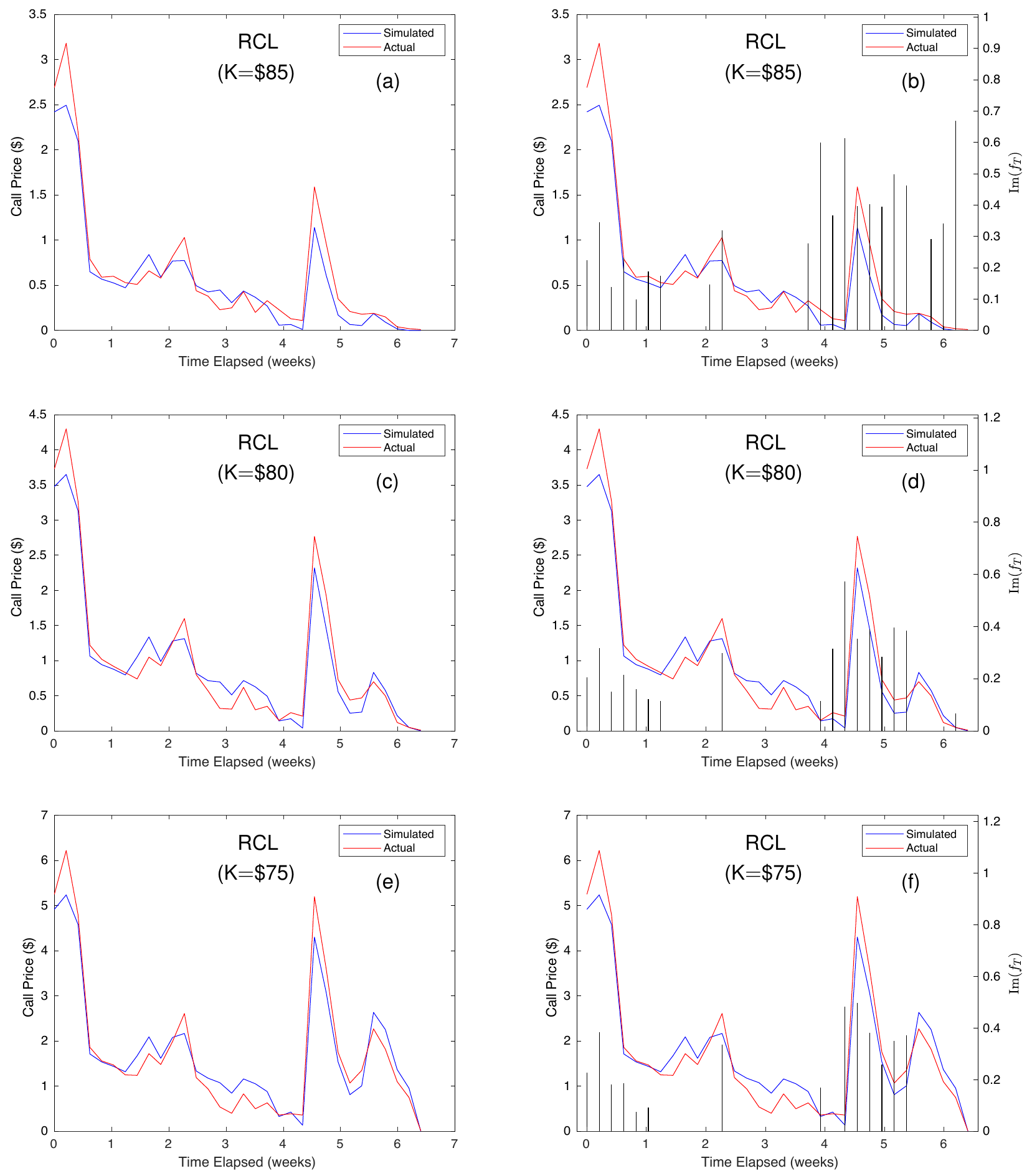}
\caption{Comparison between actual prices of a European call option (red lines) for stocks of Royal Caribbean Cruises Ltd and (a,c,e) the original Black--Scholes model based on the pricing formulae (\ref{eq2}) and (\ref{eq3}) (blue lines), and (b,c,d) the expanded Black--Scholes model based on the pricing formulae (\ref{eq4}) and (\ref{eq5}) of the expanded Black--Scholes model proposed by Segal and Segal (black vertical lines).}
\end{figure*}

\begin{figure*}[p!] 
\includegraphics[width=\textwidth]{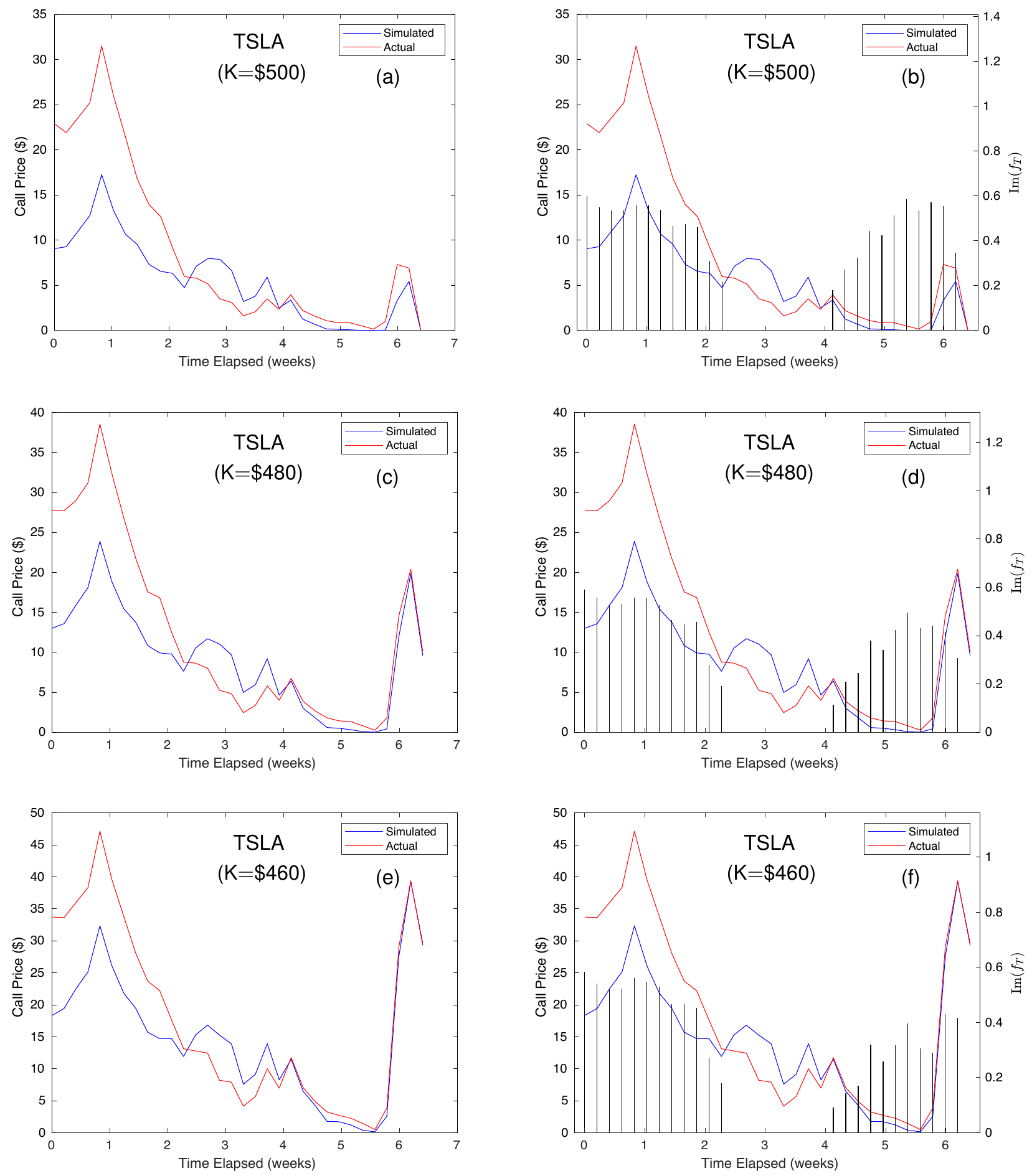}
\caption{Comparison between actual prices of a European call option (red lines) for stocks of Tesla Inc.~and (a,c,e) the original Black--Scholes model based on the pricing formulae (\ref{eq2}) and (\ref{eq3}) (blue lines), and (b,c,d) the expanded Black--Scholes model based on the pricing formulae (\ref{eq4}) and (\ref{eq5}) of the expanded Black--Scholes model proposed by Segal and Segal (black vertical lines).}
\end{figure*}

\begin{figure*}[p!] 
\includegraphics[width=\textwidth]{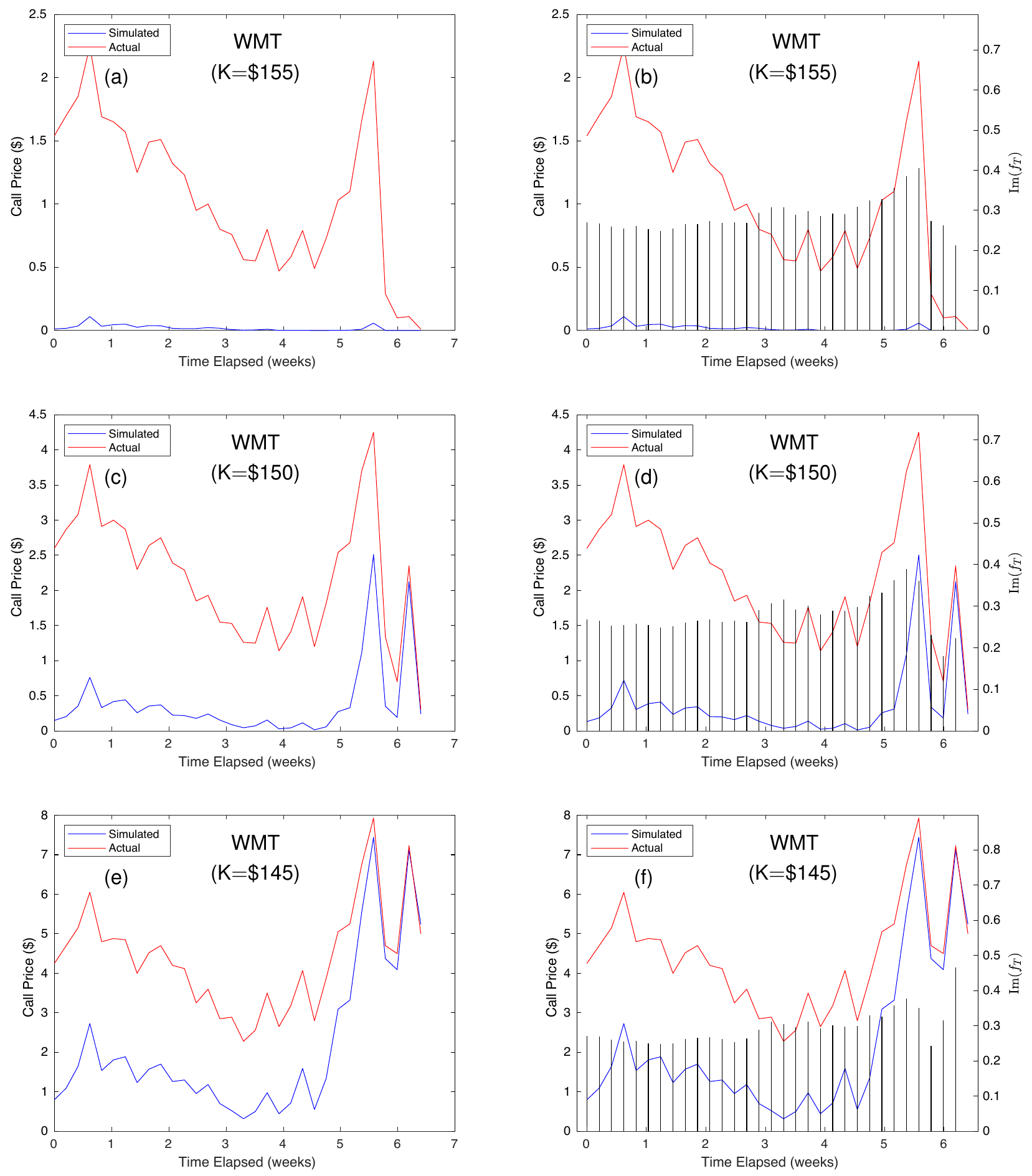}
\caption{Comparison between actual prices of a European call option (red lines) for stocks of Walmart Inc.~and (a,c,e) the original Black--Scholes model based on the pricing formulae (\ref{eq2}) and (\ref{eq3}) (blue lines), and (b,c,d) the expanded Black--Scholes model based on the pricing formulae (\ref{eq4}) and (\ref{eq5}) of the expanded Black--Scholes model proposed by Segal and Segal (black vertical lines).}
\end{figure*}

\begin{figure*}[p!] 
\includegraphics[width=\textwidth]{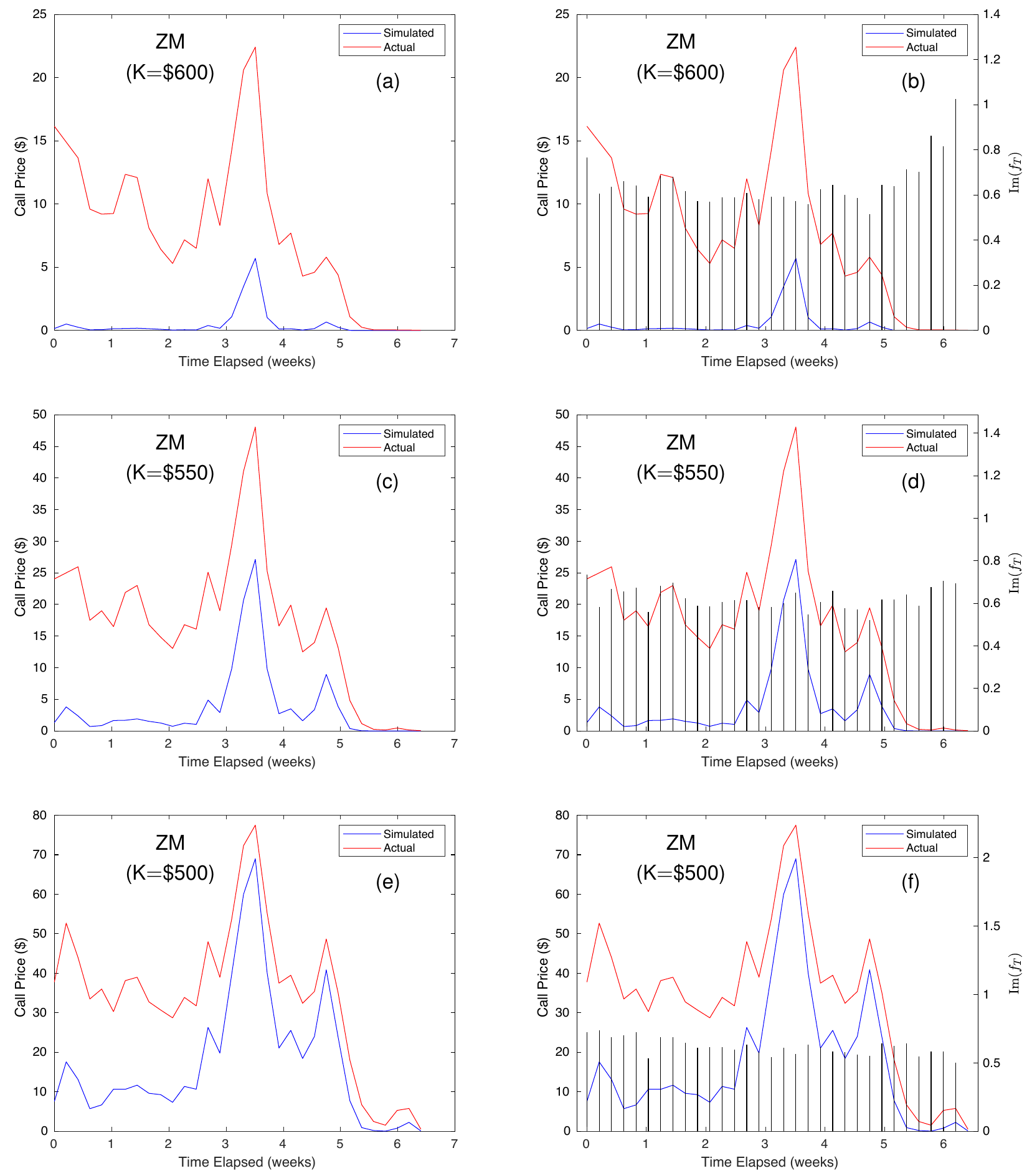}
\caption{Comparison between actual prices of a European call option (red lines) for stocks of Zoom Video Communications Inc.~and (a,c,e) the original Black--Scholes model based on the pricing formulae (\ref{eq2}) and (\ref{eq3}) (blue lines), and (b,c,d) the expanded Black--Scholes model based on the pricing formulae (\ref{eq4}) and (\ref{eq5}) of the expanded Black--Scholes model proposed by Segal and Segal (black vertical lines).}
\end{figure*}

%\bibliography{apssamp}

\bibliographystyle{elsarticle-num}

\bibliography{cas-refs}